\documentclass[reprint,aps, prfluids]{revtex4-2}

\usepackage{verbatim} 
\usepackage{graphicx}
\usepackage{dcolumn}
\usepackage{xr}
\usepackage{bm}
\usepackage{siunitx} 
\usepackage[utf8]{inputenc}
\usepackage[T1]{fontenc} 
\usepackage{amsmath}
\usepackage{float}
\usepackage{amsfonts}
\usepackage{xcolor} 
\usepackage{transparent}
\usepackage{hyperref}

\newcommand*\diff{\mathop{}\!\mathrm{d}}
\newcommand{\ro}{r_{0}}
\newcommand{\rb}{\rho_{\mathrm{b}}}
\newcommand{\rt}{\rho_{\mathrm{t}}}

\newcommand{\capco}{\hat\lambda} 

\newcommand{\Syn}{^{(\mathrm{Syn})}}
\newcommand{\targetomega}{\omega_{\mathrm{t}}}

\begin{document}

\title{Experimentally Resolving Gravity-Capillary Wave Evolution in Vessels of Unknown Boundary Conditions}

\author{Sean M. D. Gregory$^{1}$}%
\email{sean.gregory@nottingham.ac.uk}
\author{Vitor S. Barroso$^{1}$}%

\author{Silvia Schiattarella$^{2}$}

\author{Anastasios Avgoustidis$^{2}$}

\author{Silke Weinfurtner$^{1,2,3}$}

\affiliation{%
 $^{1}$School of Mathematical Sciences, University of Nottingham, University Park, Nottingham, NG7 2RD, UK
}%
\affiliation{%
 $^{2}$School of Physics and Astronomy, University of Nottingham, University Park, Nottingham, NG7 2RD, UK
}%
\affiliation{%
$^{3}$Photon Science Institute, University of Manchester, Manchester, M13 9PL, UK
}
\date{\today}

\begin{abstract}
The geometries of surface wave modes are determined by the highly nontrivial interplay of capillarity and wetting effects at the boundaries of their domain. 
Aside from idealised scenarios, this commonly leads to unknown boundary conditions, thereby hindering theoretical formulation and experimental analysis. 
To address this problem, we introduce \textit{Extracted Mode Tracking} (EMT), a data-analysis framework to obtain instantaneous amplitude and phase content of axisymmetric surface-wave modes from spatio-temporal measurements.
This approach uses unsupervised machine learning techniques to extract a basis of wave modes directly from collected data; the spatial profiles require no prior theoretical modelling, and so the issue of unknown boundary conditions is circumvented.
Time-resolved mode amplitudes are reconstructed by geometric fitting at each recorded time-step, and the success is evaluated by a spectral signal-to-noise quantifier.
Capabilities and limitations of EMT are systematically benchmarked on synthetic datasets, finding strong resilience against noise, improved accuracy over alternative methodologies, and the ability to operate with restricted domains which poses significant merit for use in experimental systems with limited measurement field-of-view.
Finally, we conduct a Faraday-wave experiment in a regime highly sensitive to boundary effects in order to further validate the method, and demonstrate the observational access to nonlinear wave-dynamics enabled by EMT.
These results establish EMT as a general tool for analysing wave mode dynamics of axially-symmetric fluid interface systems, and open pathways for quantitative studies of nonlinear mode-interactions, stability, and turbulence.
\end{abstract}

\maketitle

\renewcommand{\figurename}{\textbf{Fig.}}
\renewcommand{\thefigure}{\textbf{\arabic{figure}}}

\section{Introduction}
The properties of gravity-capillary waves, appearing upon a fluid within a confined vessel, are determined by the boundary conditions of the fluid surface at the vessel walls. 
Outside of idealised scenarios the physics of the contact line, at which the surface meets the bounding wall, is highly nontrivial~\cite{bintein:hal-04973294}; at the capillary scale, the effects of vicosity, inertia and surface interaction (through Van der Waals, electrostatic forces etc.) must all be taken into account to determine the surfaces' angle of contact and motion at the wall. 
This dynamical boundary determines the geometries of wave modes appearing on the surface.
Additionally, the capillary effects act to damp the motion of the waves and shift their frequency with a nonlinear dependence on their amplitude~\cite{cocciaro1993experimental}.
These capillary effects are highly sensitive to experimental parameters, including trace contamination and surface roughness~\cite{case1957damping} which can be very challenging to control in practice.
Without comprehensive knowledge of all chemical interaction between the fluids both above and below the surface, with each other and with the material of the vessel wall, exact boundary conditions cannot be determined.
The lack of theoretical modelling for the wave modes of a fluid system poses a significant obstacle for researchers wishing to observe their dynamics.

This problem is addressed by an ongoing avenue of research to develop progressively more sophisticated phenomenological models of the contact line motion and the subsequent damping and frequency shifts to the surface wave modes. 
The first phenomenological boundary condition to account for capillary effects was proposed by J. W. Miles~\cite{miles1967surface} who recognised that the velocity of the contact line $\left. \dot \eta \right|_{\partial \Sigma}$ was dependent on the contact angle. This is the ``Miles condition'' which, when linearised around the stationary background of the surface, $\eta_s$, can be expressed as:
\begin{equation} \label{eqn:Miles_condition_linearised}
    \left. \dot \eta \right|_{\partial \Sigma} = \capco \left. \eta' \right|_{\partial \Sigma}
\end{equation}
where $\left. \eta' \right|_{\partial \Sigma}$ is the spatial derivative of the surface excitation in the direction perpendicular to the vessel wall, and $\capco$ is a phenomenological parameter known as \textit{the capillary coefficient}. While $\capco$ was originally taken to be constant, later observations that the damping of wave modes from the dynamic boundary is dependent on their amplitude $\xi$ also~\cite{keulegan1959energy} led to the conjecture that the capillary coefficient should be amplitude dependent~\cite{cocciaro1993experimental}. 
This foundational work has been built upon over time, with increasingly sophisticated modelling being proposed to account for various effects such as the uneven motion of the contact line up and down the wall (known as capillary hysteresis)~\cite{dussan1979spreading, young1987plate}, the fluid acting to lubricate the wall~\cite{jiang2004contact}, and temporary pinning of the contact line during an oscillation cycle~\cite{bongarzone2021relaxation}.
In any such model, a capillary coefficient appears that depends on the chemical interactions between the vessel wall and the surrounding fluids.

In most scenarios, the capillary coefficient will not be known. This poses an issue for experimentation on confined surface waves in the capillary regime which can be remedied in any of three ways:
i) construct an experiment so that the Eq.~(\ref{eqn:Miles_condition_linearised}) reduces to a simplified form independent of $\capco$,
ii) experimentally determine the value of $\capco$,
iii) use an approach for which an outer boundary condition need not be specified.

Approach i) can either be achieved by pinning the contact line so that $\left. \dot \eta \right|_{\partial \Sigma}=0$ which can be taken as a Dirichlet boundary condition $\left.\eta \right|_{\partial \Sigma}=0$ without loss of generality (effectively setting $\capco =0$ in Eq.~(\ref{eqn:Miles_condition_linearised})), or by letting the contact line slip freely along the wall, in which case the surface will be perpendicular to the boundary which poses a Neumann boundary condition $\left. \eta^\prime \right|_{\partial \Sigma}=0$ (effectively $\capco \to \pm \infty$).
Methods to implement the former include increasing the surface roughness of the vessel walls (providing anchor points for the contact line to pin to, resulting in very large values of $\hat \lambda$)~\cite{de1985wetting, quere2008wetting}, or use of a vessel with a large step defect (such as a container brim) at the contact line~\cite{bechhoefer1995experimental, douady1990experimental, bongarzone2022subharmonic}. 
 A researcher aiming to construct and experiment with free-slip boundary conditions will tend to select the fluids and vessel wall material such there is very low mutual friction between contact line and wall. Alternatively, the selection can be made such that the walls can be coated by a thin film of the bottom fluid to provide a surface for a free-slipping apparent contact line to form~\cite{snoeijer2013moving, batson2013faraday}.
However, in any actual experimental implementation these limits are only approximate: capillary damping will still occur with such boundary conditions, and these limits can be difficult to approach due to various practical factors such as the presence of meniscii~\cite{douady1990experimental}.

If approach i) is infeasible or undesirable for a particular experiment, one may instead adopt approach ii): determining an empirical boundary condition by experimentation.
This may be be accomplished either by using the data collected from the experiment of interest, or by running separate precursor experiments with the express aim of ascertaining the behaviour of the fluid-vessel contact line.
Such supplementary experiments should of course use the same fluid sample and vessel wall material as the primary experiment; though an alternative detection methodology or system setup may be used if more appropriate.
Examples of experimental setups that can be used to investigate contact line behaviour include the observation of a fluid droplet spreading over a plate of the vessel material~\cite{tanner1979spreading, kavehpour2003microscopic}, and measurement of the contact line as a sample of the vessel material is inserted into the fluid at a controlled rate~\cite{inverarity1969dynamic, gutoff1982dynamic, marsh1993dynamic}. The drawback of these supplementary experiments is that the effective boundary modelling found may not match that observed in the primary experiment of interest. 
This is due to the sensitivity of the contact line to microscopic effects: minor differences between separate experimental setups can lead to different macroscopic behaviour. Controlling for trace contamination, surface roughness, and denaturing of the chemical sample over time all pose significant experimental challenges~\cite{case1957damping}.
Furthermore, various conditions may be present within the primary wave experiment that are unaccounted for in the supplementary measurements, leading to difference in the boundary conditions between the two. For example, steady and unsteady flows are expected to give rise to differing boundary conditions~\cite{jiang2004contact, miles1990capillary}, as could various other hysteresis effects~\cite{shen2024dynamic}. 
The alternative to the supplementary experimentation approach is to find a boundary condition directly from experimental data. This can be achieved by selecting an appropriate boundary condition model informed by the observed dynamics, then fitting the unknown parameters to the collected data. A few recent studies have been able to find a close fit to theoretical modelling, either by direct measurement~\cite{xia2018moving, shen2024dynamic} or comparison to numerical simulation~\cite{amberg2022detailed}. 
This direct approach is of course limited by the strength of the theoretical model adopted, and can be challenging to implement in practice~\cite{bintein2025unsteady} - especially for experimental setups not explicitly designed for a precision contact line measurement.

Finally, there is approach iii), in which the experiment is such that a description of the outer boundary condition does not need to be known. This is the case either for exploratory experiments where detailed modelling is unnecessary, or when an analysis methodology is employed that circumvents the issue of unknown boundary conditions in order to recover the evolution of the wave modes directly from the data. The most common methodologies for tracking the evolution of wave modes on a fluid surface fall into two broad categories. Time-frequency methods, such as short-time Fourier transform~\cite{grochenig2001foundations} and continuous wavelet transform~\cite{farge1992wavelet}, allow the temporal evolution of modal amplitudes to be tracked without requiring an explicit theoretical model for the spatial structure of the modes. While this model independence is a clear advantage, it comes at the cost of limited time resolution due to the inherent time-frequency uncertainty, and such approaches typically struggle to disentangle closely spaced or degenerate modes. An alternative class of methods performs a geometric decomposition of the interface at each individual time step, for example using a spatial Fourier transform, if the spatial profile of the wave modes is known to be sinusoidal, or a discrete Hankel transform if they are known to be Bessel functions~\cite{guizar2004computation,zhang2023pattern,gregory2024tracking}. These techniques place no restriction on time resolution and allow instantaneous modal amplitudes to be extracted, but they rely critically on accurate prior knowledge of the underlying wave-mode basis; mismodelling of the geometry or boundary conditions can therefore introduce systematic errors.

This paper addresses the issue of unknown boundary conditions by introduction of an analysis methodology called ``Extracted Mode Tracking'' (EMT) for tracking the time evolution of wave modes on a fluid surface.
EMT is a hybrid approach that combines the advantages of both aforementioned classes of wave mode tracking.
No intrinsic limitation is imposed on the time resolution while also avoiding the need for an assumed modal basis, instead learning the spatial structure of the modes directly from the data itself --
no specification of the contact line boundary conditions required.
In this sense, EMT entails an implementation of approach iii) by way of an unsupervised machine-learning framework.
Data informed techniques such as this are well-suited to experimental systems in which the relevant degrees of freedom are not known a priori.
By using standard linear-algebraic operations, including singular value decomposition and regularised regression, to extract both the modal shapes and their temporal evolution self-consistently, EMT enables robust mode tracking even in situations where the boundary conditions of the system are unknown.

For the purposes of demonstration we consider a fluid-fluid interface within a cylindrical vessel. However, the approach is generally applicable to any axially symmetric fluid system such as fluid surfaces in annular geometries~\cite{barroso2022primary}, or systems with non-wetting boundary conditions such as superfluids~\cite{vsvanvcara2024rotating,barroso2025digital}.
An additional utility is the technique's use on datasets of measurements with restricted access to the interface's spatial domain.
Furthermore, we provide discussion on how EMT may be generalised to fluid systems of geometries other than cylindrical.
The methodology is systematically validated on both synthetic and experimental datasets, and we showcase some physical phenomena that EMT enables the observation thereof.

\begin{figure*}[ht]
    \centering
\def\svgwidth{\linewidth}
\begingroup%
  \makeatletter%
  \providecommand\color[2][]{%
    \errmessage{(Inkscape) Color is used for the text in Inkscape, but the package 'color.sty' is not loaded}%
    \renewcommand\color[2][]{}%
  }%
  \providecommand\transparent[1]{%
    \errmessage{(Inkscape) Transparency is used (non-zero) for the text in Inkscape, but the package 'transparent.sty' is not loaded}%
    \renewcommand\transparent[1]{}%
  }%
  \providecommand\rotatebox[2]{#2}%
  \newcommand*\fsize{\dimexpr\f@size pt\relax}%
  \newcommand*\lineheight[1]{\fontsize{\fsize}{#1\fsize}\selectfont}%
  \ifx\svgwidth\undefined%
    \setlength{\unitlength}{750bp}%
    \ifx\svgscale\undefined%
      \relax%
    \else%
      \setlength{\unitlength}{\unitlength * \real{\svgscale}}%
    \fi%
  \else%
    \setlength{\unitlength}{\svgwidth}%
  \fi%
  \global\let\svgwidth\undefined%
  \global\let\svgscale\undefined%
  \makeatother%
  \begin{picture}(1,0.25)%
    \lineheight{1}%
    \setlength\tabcolsep{0pt}%
    \put(0,0){\includegraphics[width=\unitlength,page=1]{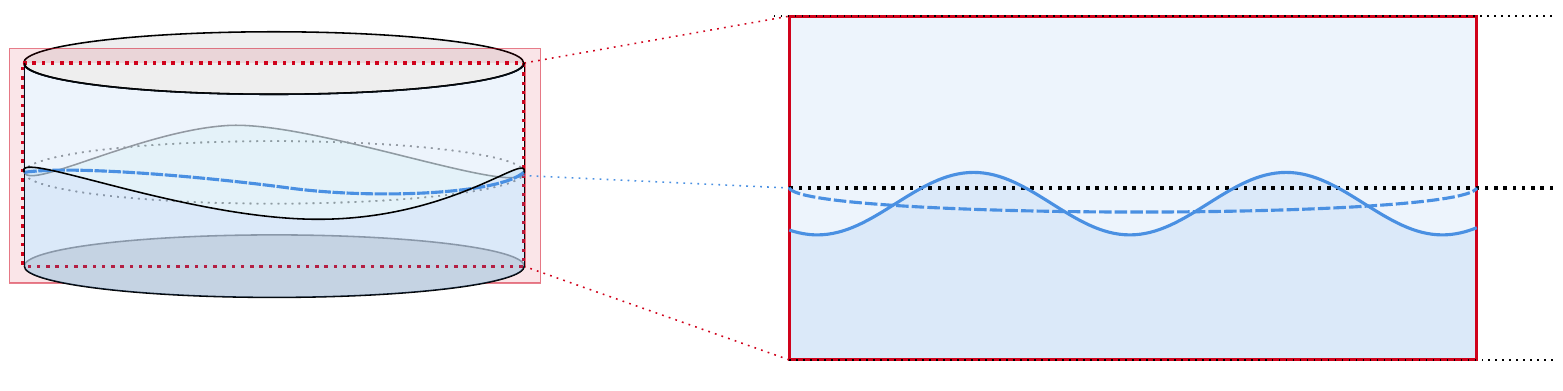}}%
    \put(0.94657501,0.13407654){\color[rgb]{0,0,0}\transparent{0.82999998}\makebox(0,0)[lt]{\lineheight{1.25}\smash{\begin{tabular}[t]{l}$z=0$\end{tabular}}}}%
    \put(0.94558606,0.02426567){\color[rgb]{0,0,0}\transparent{0.82999998}\makebox(0,0)[lt]{\lineheight{1.25}\smash{\begin{tabular}[t]{l}$z=-h_{\mathrm{b}}$\end{tabular}}}}%
    \put(0.94556329,0.22398296){\color[rgb]{0,0,0}\transparent{0.82999998}\makebox(0,0)[lt]{\lineheight{1.25}\smash{\begin{tabular}[t]{l}$z=h_{\mathrm{t}}$\end{tabular}}}}%
    \put(0.60999725,0.09790044){\color[rgb]{0,0,0}\transparent{0.82999998}\makebox(0,0)[lt]{\lineheight{1.25}\smash{\begin{tabular}[t]{l}$\eta_{\mathrm{s}}$\end{tabular}}}}%
    \put(0.61278131,0.14703211){\color[rgb]{0,0,0}\transparent{0.82999998}\makebox(0,0)[lt]{\lineheight{1.25}\smash{\begin{tabular}[t]{l}$\eta$\end{tabular}}}}%
  \end{picture}%
\endgroup%

\caption{\textbf{Cross section of the fluid system} 
The dynamical object of interest is the interfacial surface that forms between two immiscible fluids of differing densities. 
The horizontal domain of the interface is a disk bounded by impermeable horizontal walls at radius $r=r_0$.
The interface is composed of dynamic fluctuations $\eta(t,r,\theta)$ around a static background $\eta_{\mathrm{s}}(r)$ which may have curvature as it approaches $r_0$ due to the formation of a meniscus. We define $z=0$ at the static contact line $\eta_{\mathrm{s}}(r_0)$, which is at distances $h_\mathrm{b}$, $h_\mathrm{t}$ from the horizontal floor and ceiling respectively.
}
\label{fig:Diagram_of_cross_section}
\end{figure*}
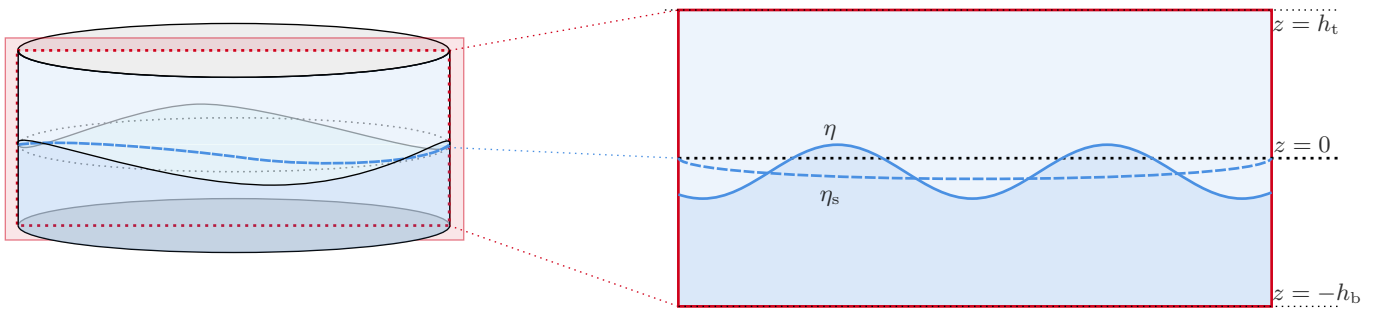

The paper is organized as follows. Sect.~\ref{Fluid_system_section} establishes the physical system in consideration. In Sect.~\ref{Govering_eqs_subsection} the governing equations for a fluid surface are provided and our formalism for the description of wave modes defined for the specific geometry we take. Sect.~\ref{Faraday_wave_subsect} details the physics expected of a Faraday wave experiment -- the dynamical scenario under consideration. In Sect.~\ref{experimental_setup_subsection} we specify the details of our own laboratory experiment conducted for the purposes of demonstrating our novel methodology. 
Sect.~\ref{Mode_tracking_section} presents the EMT protocol. We provide a step-by-step explanation of the procedure split into the extraction of the wave mode basis in Sect.~\ref{Mode_extraction_subsubsect}, and the amplitude tracking through time in Sect.~\ref{Mode_tracking_section}.
Sect.~\ref{Generalisation_subsect} concludes the section with a discussion of how EMT may be generalised for application on other experimental systems with arbitrarily shaped vessels. 
Sect.~\ref{Validation} serves to thoroughly validate the Method. In Sect.~\ref{EMT_benchmarking},
synthetic datasets are prepared for the purposes of benchmarking the capabilities of EMT and comparing against alternative methodologies. In Sect.~\ref{Experimental_results_subsection}, we present the results from applying EMT on real experimental data documenting the success and highlighting the capabilities.

\section{The Fluid system}\label{Fluid_system_section}
\subsection{Two-fluid system} \label{Govering_eqs_subsection}

The fluid system we consider is composed of two stratified fluid layers contained within a cylindrical vessel. The ``bottom fluid'' forming the bottom layer has density $\rb$ and kinematic viscosity $\nu_{\mathrm{b}}$, and the ``top fluid'' has density $\rho_{\mathrm{t}}$ and kinematic viscosity $\nu_{\mathrm{t}}$. The fluids are immiscible with $\rb>\rt$, and so are separated by an interfacial surface $z-\Gamma(t,r,\theta)=0$ with surface tension $\sigma_{\mathrm{st}}$. We separate this surface into dynamical fluctuations $\eta$ around a static background $\eta_{\mathrm{s}}$, i.e. $\Gamma(t,r,\theta)=\eta_{\mathrm{s}}(r)+ \eta(t,r,\theta)$. Note that $\eta_{\mathrm{s}}$ is only a function of $r$ due to the axial symmetry of the system, and may generally be curved due to the formation of a meniscus. Defining $z=0$ at the resting contact line $\eta_{\mathrm{s}}(\ro)=0$ as illustrated in Figure~\ref{fig:Diagram_of_cross_section}, the impermeable floor and ceiling of the vessel are located at $z=-h_{\mathrm{b}}$ and $z=h_{\mathrm{t}}$ respectively, and the interface is bounded by an impermeable vertical wall at $r=\ro$. 
It should be noted that all of this formulation can easily be extended to an annular geometry by introducing a second non-zero boundary on the radial domain.

The fluids (labelled $j=\{\mathrm{b},\mathrm{t}\}$) are incompressible, and so their respective velocity potentials $\phi_j (t,r,\theta,z)$ are governed by the Navier-Stokes equations:
\begin{subequations}
\label{eqn:Navier_stokes}
\begin{align}
\rho_j \left[\frac{\partial}{\partial t} + \nabla \phi_j \cdot \nabla \right] \nabla \phi_j &=\nabla \cdot \overleftrightarrow{\pi}_j + \Vec{f}_j, \label{eqn:NS_eom} \\
\nabla^2 \phi_j &=0, \label{eqn:NS_incompressibility} 
\end{align}
\end{subequations}
where $\overleftrightarrow{\pi}_j$ and $\Vec{f}_j$ are the stress tensor and the external force density on fluid $j$. 
The flow is irrotational, and so the requirement that fluid particles are advected by the flow (i.e. the material derivative of $\Gamma$ vanishes) leads to the following kinematic boundary condition at the interface:
\begin{equation}
    \frac{\partial \eta}{\partial t} =\frac{\partial \phi_j}{\partial z}  -  \nabla_{\perp} \eta \cdot  \nabla \phi_j- \frac{\partial\phi_j}{\partial r} \frac{\partial\eta_\mathrm{s}}{\partial r}, \quad \text{at } z= \Gamma,
\end{equation}
which relates the velocity potentials to the interface height.

While the boundary conditions are unknown for the surface $\Gamma$,
the impermeability of the vessel's walls, floor, and ceiling impose the following boundary conditions upon the velocity potential of each fluid:
\begin{equation} \label{eqn:impermeability_bcs}
    \left.\frac{\partial \phi_j}{\partial r}\right|_{\ro} =
\left.\frac{\partial \phi_j}{\partial z}\right|_{z=-h_{\mathrm{b}}}
= \left.\frac{\partial \phi_j}{\partial z}\right|_{z=h_{\mathrm{t}}}  =0.
\end{equation}

We describe the interface fluctuations as a linear superposition of wave modes, each consisting of a dimensionless characteristic spatial profile $\Psi_{a}(r,\theta)$ and a time dependent amplitude $\xi_{a}(t)$ with dimensions of length:
\begin{equation} \label{eqn:general_decomposition}
    \eta(t, \Vec{x}) = \sum_{a}  \xi_{a}(t) \Psi_{a}(r,\theta).
\end{equation}
This basis is discretised in space due to the vessel's finite size.
Each amplitude oscillates with one specific frequency at an envelope amplitude $\overline{\xi}_a(t)$ that varies on timescales slower than the oscillation:
\begin{equation} \label{eqn:ximn_time_dependence}
    \xi_{a} (t) =\frac{1}{\sqrt{2}} e^{i \omega t} \overline{\xi}_{a}(t).
\end{equation}
The envelope amplitude is complex, with its complex phase describing the angular and temporal phase of the mode, and its magnitude $|\overline{\xi}_a|$ equal to the root-mean-squared height of the mode over one oscillation cycle.

The spatial profiles $\Psi_a$ are determined by the geometry and boundary conditions of $\xi$. The advantage to a cylindrical geometry is that $\Gamma$ must be periodic in the azimuthal dimension, and so the mode spatial profiles can always be decomposed as:
\begin{equation} \label{eqn:Psi_cylinder}
    \Psi_{m,\omega} (r,\theta) = R_{m,\omega}(r) e^{i m \theta}.
\end{equation}
Note that we have now adopted the label $a =(m,\omega)$ as the azimuthal number $m$ and frequency $\omega$ are sufficient to uniquely specify a wave mode.
Since the height field $\eta$ must be real valued, the azimuthal number $m$ is summed over positive and negative values in Eq.~(\ref{eqn:general_decomposition}), with the relation $\xi_{-m,\omega} = \xi_{m,\omega}^*$ and $R_{-m,\omega} = R_{m,\omega}^*$. Standing waves occur whenever there is the symmetry $\overline{\xi}_{m,\omega}=\overline{\xi}_{m,-\omega}$, $R_{m,\omega}=R_{m,-\omega}$.

The height displacement boundary condition at $r=r_0$ is considered to be entirely unknown, and so the radial functions $R_{m,\omega}$ cannot be determined. We adopt the convention that these are complex dimensionless functions $R_{m,\omega}:\mathbb{R} \to \mathbb{C}$ on the domain $r\in [0,r_0]$ that are each normalised such that:
\begin{equation} \label{eqn:R_normalisation}
    \langle |R_{m,\omega}|^2 \rangle_r \equiv \frac{2}{r_0^2} \int_0^{r_0} r |R_{m,\omega}|^2 \diff r = 1,
\end{equation}
where $\langle \cdot \rangle_r$ denotes an average over the radial domain.

The height field decomposition of Eq.~(\ref{eqn:general_decomposition}) implicitly assumes that each wave mode is separable into a spatially dependent component $\Psi_a(r, \theta)$ and a single-frequency time dependent component $\xi_a(t)$. While this description is suitable for most dynamical scenarios, spatial profiles $\Psi_a$ may in principle possess non-separable time dependence as a result of strong nonlinear effects~\cite{cocciaro1993experimental, bongarzone2021relaxation}.
For the purposes of the EMT method, the effectiveness of the model Eq.~(\ref{eqn:general_decomposition}) is assumed. 
Breakdown of this assumption can be identified during the mode extraction stage of the EMT procedure, as will be discussed later in Sect.~\ref{Mode_extraction_subsubsect}.

\begin{figure*}[ht]
\centering
\includegraphics[width=\linewidth]{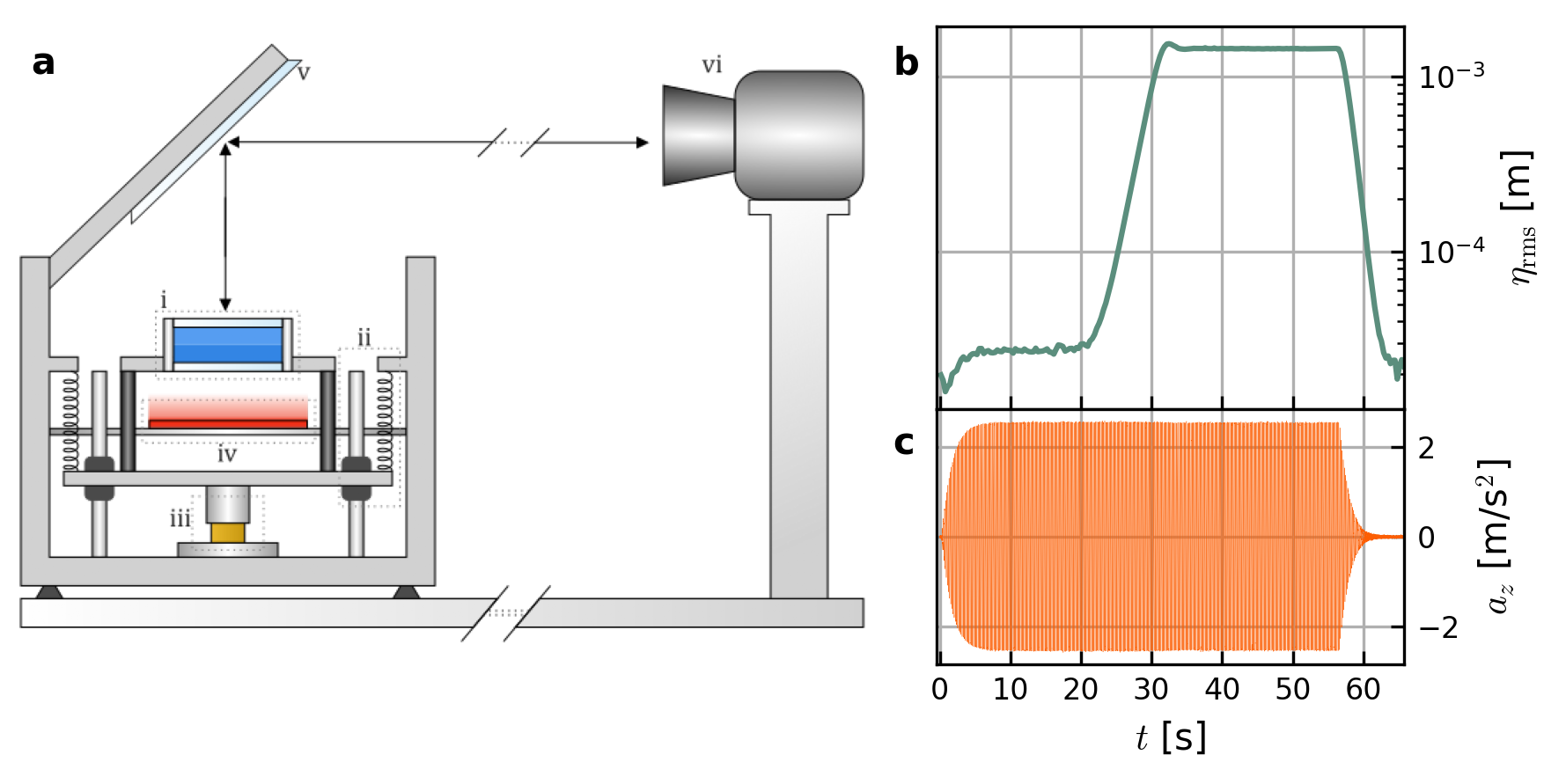}
\caption{\textbf{Faraday wave experiment} 
Panel \textbf{a} depicts our experimental setup.
A cylindrical vessel (i) containing the two-fluid system is mounted to a bespoke shaking platform. The platform is suspended by springs and glides freely in the $z$ direction along pneumatic air bearings (ii). A voice coil actuator (iii) provides a sinusoidal acceleration at a frequency
$\omega_{\mathrm{d}}=6.2(2\pi)\mathrm{Hz}$ with amplitude $|a_z|=\SI{2.53(4)}{\meter \second^{-2}}$. This is measured via a platform-mounted accelerometer (presented in panel \textbf{c} for one experimental run). Light from a backlit chequerboard pattern (iv) passes through the fluid-fluid interface where it reflects off a mirror (v) and passes to the high-speed camera (vi). This recorded data can be used to reconstruct the interface fluctuation height $\eta(t,r,\theta)$ via Fourier Transform Profilometry. 
Panel \textbf{b} displays $\eta_{\mathrm{rms}}$:
the root-mean-square of $\eta$ over the spatial domain and time intervals of $4\pi/\omega_{\mathrm{d}}$, over the course of the same experiment as that presented in panel \textbf{c}. One observes an exponential growth of $\eta_{\mathrm{rms}}$ above the initial noise level at around $\SI{20}{\second}$ after the actuator starts oscillating. Around 12 seconds later, $\eta_{\mathrm{rms}}$ stabilises as the system enters a nonequilibrium steady-state. The steady-state persists until the driving is switched of and $\eta_{\mathrm{rms}}$ rapidly decays back to equilibrium.
}
\label{fig:Faraday_wave_experiment}
\end{figure*}

\subsection{Faraday wave evolution} \label{Faraday_wave_subsect}

As a demonstration of EMT and the physics it reveals, we consider a Faraday wave experiment~\cite{Faraday}. Mechanical oscillation of the fluid vessel is used as a means to source the growth of some wave modes, which subsequently excite more wave modes through nonlinear wave-mixing. Eventually, a nonequilibrium steady state is attained at which all excited modes oscillate with constant envelop amplitudes. Once the mechanical oscillation is switched off, the interface decays to its equilibrium configuration.
This section focuses on the theoretical underpinning, while Sect.~\ref{experimental_setup_subsection} will detail how the scenario was experimentally achieved, as seen in Fig.~\ref{fig:Faraday_wave_experiment}\textbf{b}, by use of a bespoke shaking platform set-up (Fig.~\ref{fig:Faraday_wave_experiment}\textbf{a}).

The scenario was chosen due to its convenience for the demonstration of EMT. This is because data collected during the nonequilibrium steady-state is ideal for the purposes of extracting mode spatial profiles. This will become clear as the procedure is explained in Sect.~\ref{Mode_extraction_subsubsect}, but we note that the two beneficial qualities of the steady-state are: a) all excited modes simultaneously have relatively large and constant amplitudes, and so the signal is clear, b) continuous driving lets one collect an arbitrarily high volume of data for a single experimental run, which improves accuracy of mode extraction (see Sect.\ref{EMT_benchmarking} Fig.\ref{fig:EVR_vs_Nt}). However, it should be stressed that the EMT method is applicable to any fluid wave experiment with spatio-temporal measurement of the surface. If one wishes to apply the EMT method to an experiment with more limited data (e.g. the dynamics do not include a steady-state, the experimental runs are very brief), one could consider remedying this problem by aggregating data from an ensemble of repetitions, or even performing a precursor Faraday wave experiment within the same system if the experimental setup allows for this.

In a Faraday wave experiment, a periodic forcing is applied to the fluid surface at some driving frequency $\omega_\mathrm{d}$. To reduce mechanical noise and preserve axial symmetry, this is best done by a controlled vertical oscillation of the fluid vessel. This acceleration $a_z(t)\propto\sin(\omega_\mathrm{d}t)$ enters into the external force term equation of motion Eq.~(\ref{eqn:NS_eom}) as $\Vec{f}_j = - \nabla(\rho_j z (g + a_z ))$, where $g$ is gravitational acceleration.
At linear order, expanding $\eta$ into the modal basis Eq.~(\ref{eqn:general_decomposition}) with this forcing leads to the following equation of motion for each mode amplitude:
\begin{equation} \label{eqn:Mathieu}
    \frac{\partial^2 \xi_a}{\partial t^2} + 2\gamma_a \frac{\partial \xi_a}{\partial t}  + \left[\omega_{a,0}^2 + \epsilon \sin (\omega_{\mathrm{d}} t)\right] \xi_a  = 0,
\end{equation}
where $\gamma_a$ is the viscous damping, $\omega_{a,0}$ is the unforced dispersion frequency, and $\epsilon$ is a driving strength parameter. This can be recognised as the damped Mathieu's equation, solutions of which have been extensively studied~\cite{magnus2013hill, kovacic2018mathieu}. Unstable solutions exist at frequencies within bands (Floquet tongues) around half-integer multiples of $\omega_\mathrm{d}$ and above a forcing strength, known as the Faraday threshold, needed to overcome the viscous damping. 
Damping rates, and hence Faraday thresholds, are mode dependent: increasing with frequency and decreasing with lengthscale~\cite{faraday1831xvii}. 
Wave modes with frequencies $\omega$ and damping rates $\gamma_{m,\omega}$ falling within Floquet tongues are unstable, and so their envelope amplitudes are amplified exponentially $\overline{\xi}_{m,\omega} \propto \exp{(\lambda_{m,\omega} t)}$. These unstable modes are known as \textit{Faraday waves}. Floquet theory allows for numerical calculation of the growth rates $\lambda_{m,\omega}$~\cite{kovacic2018mathieu}.

Let us consider a low frequency drive such that energy is injected into the system at a length scale on the same order as its size $\omega_{\mathrm{d}} /\sqrt{g_0 \mathrm{At}} \sim 2 r_0$ (where $\mathrm{At}= (\rb - \rt)/(\rb + \rt)$ is the Atwood number). Since the modal basis is discretised, with sparser separation at low frequencies, the system can be tailored such that only one $m=m_\mathrm{p}$ mode of frequency $\omega_\mathrm{p} \simeq \omega_{\mathrm{d}}/2$, herein referred to as the primary, falls within the sub-harmonic Floquet tongue. Let us assume also that the driving strength has been adjusted such that wave modes at higher harmonic tongues lie below the Faraday threshold (which can readily be accomplished since, for wavelength scales smaller than the fluid depth, the damping rate increases with frequency~\cite{case1957damping,herreman2019perturbation}). The primary mode is therefore the only Faraday wave appearing on the interface.

At early times, while the amplitude $\xi_{\mathrm{p}}\equiv\xi_{m_\mathrm{p},\omega_\mathrm{p}}$ of the primary mode is small, the interface dynamics are effectively captured by the linear Mathieu equation ~(\ref{eqn:Mathieu}). However, as $\xi_{\mathrm{p}}$ continues to grow the nonlinear terms of Eq.~(\ref{eqn:NS_eom}) become increasingly relevant. This leads to wave mixing interactions between multiple modes. While the exact strengths of specific mode-mode interactions can be calculated from a recently verified Lagrangian formalism~\cite{miles1984nonlinear,Barroso_2023,gregory2024tracking}, for our purposes it is sufficient to state that all such interactions conserve both the azimuthal numbers and frequencies of the modes involved (i.e. if three modes of azimuthal numbers $m_1$, $m_2$, $m_3$ and frequencies $\omega_{1}$, $\omega_{2}$, $\omega_{3}$ interact, it must be the case that $|m_3| =  |m_1 \pm m_2|$ and $|\omega_{3}| =  |\omega_{1} \pm \omega_{2}|$). Through these wave-mixing interactions, the primary mode sources the growth of a few other modes, known as secondary, and as these grow larger they interact to source the growth of many more wave modes on the interface. Since there is only one primary mode, of azimuthal number $m_\mathrm{p}$ and frequency $\omega_{\mathrm{p}}$, and these quantities are conserved in nonlinear interaction processes, all other modes excited must have $m = n_m m_\mathrm{p}$, $\omega = n_\omega \omega_\mathrm{p}$, where $n_m$, $n_\omega$ are either both even or both odd integers.

Of course the exponential amplification of these modes cannot last indefinitely; eventually the growth amplitudes are arrested by nonlinear effects and the system stabilises at a nonequilibrium fixed point. So long as the driving oscillation remains constant, the steady-state persists. Once the driving is stopped, damping processes will cause the waves to decay and the interface will relax back to its stationary configuration $\Gamma=\eta_s$.

\subsection{Experimental setup} \label{experimental_setup_subsection}

We validate our proposed methodology by applying it to a dataset collected from a Faraday-wave experiment as detailed in Sect.~\ref{Faraday_wave_subsect}. The fluid dynamical setup used was a two-fluid system within a cylindrical vessel, as detailed in Sect.~\ref{Govering_eqs_subsection}.
By using a confined two-fluid configuration instead of a free surface, the upper immiscible layer shields the working fluid from direct exposure to air, suppressing evaporation between experimental repetitions. This mitigates slow drifts in fluid depth and material parameters, thereby improving the repeatability of the experiment and enabling more reliable ensemble statistics.
The bottom fluid is an aqueous solution of potassium carbonate and distilled water with density $\rho_{\mathrm{b}}=\SI{1288.7\pm2.3}{\kilogram\per\meter^3}$ and kinematic viscosity $\nu_{\mathrm{b}}=\SI{2.35(2)}{\milli\meter^2\per\second}$, while the top fluid is an organic solution of ethanol and distilled water with density $\rho_{\mathrm{t}}=\SI{920.6\pm2.3}{\kilogram\per\meter^3}$ and viscosity $\nu_{\mathrm{t}}=\SI{3.40(2)}{\milli\meter^2\per\second}$.
The immiscibility of the fluids causes them to form a stratified biphasic liquid solution with the two layers separated by a surface of interfacial tension $\sigma=\SI{3.0(5)}{\milli \newton \per \meter}$. Equal volumes of both fluids are used, and so we define $z=0$ as an equidistant cross-section of the $\SI{30}{\milli\meter}$ tall vessel, i.e. $h_\mathrm{b}=h_\mathrm{t}=\SI{15}{\milli\meter}$.

The cylindrical vessel itself is composed of a transparent glass ceiling and floor, with rigid side walls made of machined nylon at $\ro=\SI{40}{\milli\meter}$. A bespoke shaking platform, on which the vessel is mounted, is suspended by springs from a base securely mounted on an optical table to reduce vibrational noise. The platform is sinusoidally driven by a voice coil actuator, while vertically gliding along pneumatic air bearings to reduce off-axis oscillations, providing a nearly entirely vertical oscillating acceleration to the vessel.
The precise control of the forcing acceleration enabled by shaking platform is essential for the strong repeatability observed across experimental runs.
The typical horizontal to vertical acceleration ratio is $|a_\perp|/|a_z|=\num{4.3e-3}$ for the datasets presented here.
The shaking platform operates at $\omega_{\mathrm{d}}=6.2(2\pi)\mathrm{Hz}$ with a driving acceleration of $|a_z|=\SI{2.53(4)}{\meter \second^{-2}}$. 
A schematic of the experimental set-up is depicted in Figure~\ref{fig:Faraday_wave_experiment}\textbf{a}, and the (accelerometer recorded) vertical acceleration profile is displayed in Figure~\ref{fig:Faraday_wave_experiment}\textbf{c}.
The low-frequency periodic forcing sources the growth of wave modes at lengthscales comparable to the vessel diameter. It is exactly these wave modes of the system which are most susceptible to the outer boundary conditions so are most relevant to our area of study.

Our choice of chemical solutions for the two-fluid system enables the tuning of physical properties in each layer and their interfacial tension by varying the concentrations of each component~\cite{salabat2007liquidliquid}. Due to the alkali nature of the aqueous solution and the strong solvent capabilities of the organic one, the resulting fluid layers are highly chemically reactive to most metals and some polymers. These properties restrict the available materials for manufacturing the side walls of the cylindrical vessel. A natural choice would be glass, which is chemically stable with the solutions and allows the two-fluid interface to sit almost exactly perpendicular to the walls, i.e., a negligible meniscus forms. However, for our purposes here, we opted for a polymer with a rougher finish (e.g., compared to glass), namely nylon. 

The machined nylon cylindrical walls are chemically stable enough not to degrade over time, while providing a steeper meniscus between the two-fluid interface and the walls. The latter creates non-trivial wetting conditions, facilitating the study of unknown boundary conditions on the normal modes of the confined interface. To apply our methodology for extracting the dynamical and spatial evolutions of surface modes, we require the spatially and temporally resolved profile of the two-fluid interface. In the experimental implementation discussed here, we opted for a detection scheme that yields the full-field reconstructions of the surface. However, we note that our analysis methods also apply to spatial data with a restricted field of view, as will be demonstrated in Sect.~\ref{EMT_benchmarking}.

Waves on the two-fluid interface are detected using modified Schlieren imaging with a periodic backdrop~\cite{Moisy2009ASS, Wildeman2018Real-timeBackdrop}. A backlit chequerboard pattern is placed underneath the oscillating vessel and statically mounted to the base. At 2 metres away from the pattern, a high-speed camera acquires images of the system through telecentric lenses. Interfacial waves or disturbances appear as local deformations to the checkerboard pattern. They can be related to the gradient of the interface height, which can be retrieved numerically through Fourier Transform Profilometry and inverted to reconstruct the complete profile of the interface~\cite{Wildeman2018Real-timeBackdrop}. 
Our implementation of this method yields surface-height resolution of around one micron.

\section{Extracted mode tracking}\label{Mode_tracking_section}
This section details ``Extracted Mode tracking'' (EMT): a novel methodology for resolving individual wave modes of the fluid-fluid interface (introduced in Sect.~\ref{Fluid_system_section}), and tracking their complete evolution through time.
The wave modes comprising the basis used in this approach are extracted directly from experimental data. This has the advantage of circumventing the need to theoretically model the form of the modes' spatial profiles $\Psi_{m,\omega}$, which is often infeasible in non-trivial geometries/ boundary conditions.

\begin{figure*}[ht]
\centering
\includegraphics[width=\linewidth]{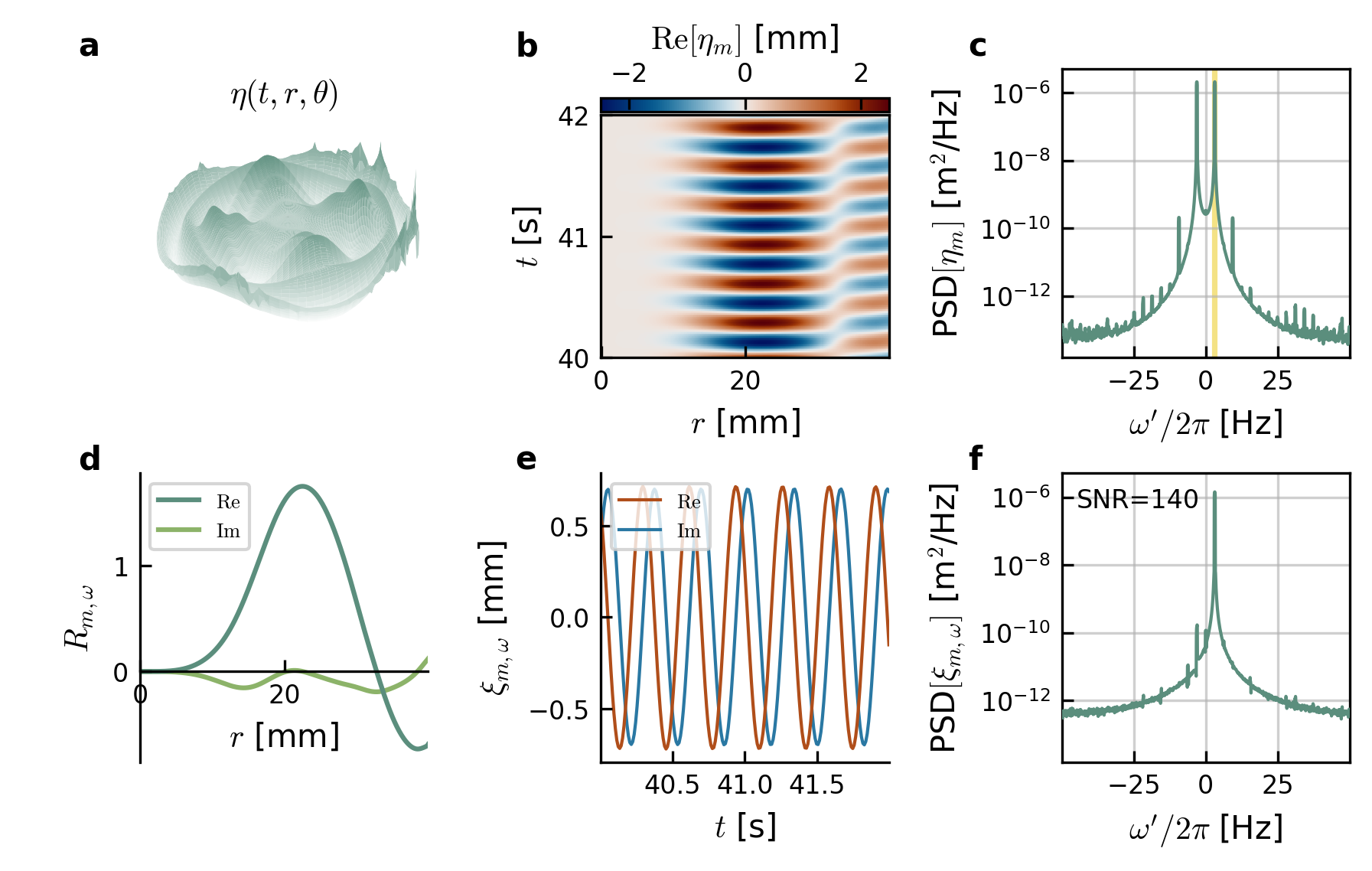}
\caption{\textbf{Mode extraction} 
The wave mode extraction protocol is illustrated for the example of the $m=4, \omega = 3.1(2\pi)\mathrm{Hz}$ mode in data from the experiment detailed in Section~\ref{experimental_setup_subsection}. 
Data is collected on the evolution of $\eta(t,r,\theta)$, a snapshot of which is displayed in panel \textbf{a}. Fourier transforming over $\theta$, and fixing an azimuthal number, one obtains $\eta_m (t,r)$ - the real component of which is displayed in panel \textbf{b}. The PSD of this quantity is calculated during the steady state (panel \textbf{c}). Frequencies at which peaks are found in the PSD correspond to the frequencies at which modes are oscillating. 
Data of $\Tilde{\eta}_m(\omega',r)$ is band-pass filtered in a narrow frequency domain $\omega'\in[\omega - \mathrm{ENBW}, \omega + \mathrm{ENBW}]$, shaded in yellow, then passed to a truncated SVD algorithm to find a principal component proportional to the radial function of the wave mode. The radial function for $\omega = 3.1(2\pi)\mathrm{Hz}$ was extracted with an EVR of 1 to 6 significant figures, and is displayed in panel \textbf{d}. After extracting the radial profiles for each other mode found in $\mathrm{PSD}[\eta_m]$, one can track the evolution of each by a geometric fitting at each recorded time frame. The complex amplitudes $\xi_{m,\omega}$ obtained have the same time resolution as the original recording, and panel \textbf{e} displays this for our example mode. To confirm how successfully the mode was tracked, one can examine the PSD of the obtained amplitude. The majority of the signal should be present at $\omega'=\omega$, which is confirmed visually in panel \textbf{f}, and quantitively by the signal-to-noise ratio of $\mathrm{SNR}=140$ (to 3 significant figures) as defined in Eq.~(\ref{eqn:SNR_def}).
}
\label{fig:EMT_overview}
\end{figure*}

\subsection{Mode extraction} \label{Mode_extraction_subsubsect}

The first task is to extract the spatial profiles of the wave modes from the experimental data.
Let us assume that one has access to the full experimental data of the observable $\eta$ over a period of time and a region of space (which does not necessarily have to be the entire bounded domain, as will be discussed in Sect.~\ref{EMT_benchmarking}) throughout the course of a Faraday wave experiment, as detailed in Sect.~\ref{Faraday_wave_subsect}. 
The data is first separated by azimuthal number $m$ by performing a Fourier transform over angle. This is achieved numerically via a Fast Fourier Transform (FFT) algorithm to obtain data as function $\eta_{m}(t,r)$ of time and radius after fixing an azimuthal number $m$,
\begin{equation} \label{eqn:etam_exp}
    \left.\eta_{m} (t,r)  = \mathcal{F}_{\theta}[\eta (t,\theta,r)]\right|_{m}, 
\end{equation}
where $\mathcal{F}_{x}$ denotes a Fourier transform over the quantity $x$.
This is equivalent to the sum of all different (discrete) frequency modes that have the same azimuthal number $m$: $\eta_{m}  =\sum_{\omega} \xi_{m,\omega} R_{m,\omega}$.

Next, we identify the frequencies at which $m$-modes are exicted. It is best to use a strong and steady signal for this, and so we inspect a time interval $t^{\mathrm{s}}=[t^{\mathrm{s}}_{\mathrm{i}},t^{\mathrm{s}}_{\mathrm{f}}]$ during the late times, once the system has attained a steady state and envelope amplitudes have large, constant values $\overline{\xi}_{m,\omega}^{\mathrm{s}} = \overline{\xi}_{m,\omega}(t\in t_{\mathrm{s}})$. 
This signal Fourier transformed to the frequency $\omega^\prime$ domain is:
\begin{equation} \label{eqn:FFT_over_time}
    \Tilde{\eta}_m(\omega^\prime,r) =  \mathcal{F}_{t}[\eta_m (t\in t^{\mathrm{s}},r )].
\end{equation}
From this quantity a power spectral density (PSD) is calculated for $\eta_m$ during the time interval $t^{\mathrm{s}}$ (averaging out the radial dependence):
\begin{equation} \label{eqn:PSD_eta_m}
    \mathrm{PSD}[\eta_m] (\omega^\prime)\propto \langle | \Tilde{\eta}_m(\omega^\prime,r)|^2 \rangle_r.
\end{equation}
Since the data will contain noise from the experimental measurement and numerical calculation error, it is advisable to use a periodogram to estimate the PSD rather than a standard FFT.
We chose a multitapering approach for this purpose, due to its ability to simultaneously preserve frequency resolution while providing resilience against noise~\cite{percival1993spectral}.

Since mode amplitudes evolve according to Eq.~(\ref{eqn:ximn_time_dependence}), excited modes will appear as peaks proportional to $|\overline{\xi}_{m,\omega}|^2$ in the PSD.
These peaks occur at their oscillation frequency but are broadened by the Equivalent Noise Bandwidth (ENBW) of the periodogram used for calculating the PSD.
The ENBW is set by the value of time-half bandwidth product, NW, used for the multitapering periodogram. The NW is the tuning parameter of the multitaper method: with increasing values there is a trade-off between better noise suppression and poorer frequency resolution. Typically values of $2\leq \text{NW} \leq4$ are taken for conventional data analysis purposes~\cite{percival1993spectral}; for the experimental results presented within this paper, $\text{NW}=3$ has been selected as a reasonable middle-ground.
Knowing the ENBW to correspond to the expected peak width, a standard peak finding algorithm can identify the frequencies $\omega_{m,i}$ ($i=0,1,2,\dots$), of significant spectral peaks above the background noise, up to a threshold set by the resolution of the data.

Finally, for each $m$, we must extract the radial profiles for the modes in the set of identified frequencies 
$\omega_{m,\{i\}}$, where we denote a set of values by $\{ \cdot\}$ around the varying index. 
Radial data $\Tilde\eta_m$ is isolated around each frequency $\omega_{m,i}$ by use of a bandpass filter with a width set by the periodogram ENBW. Through a truncated singular value decomposition (SVD)~\cite{manning2008matrix} on this isolated data, one can identify the principal component as proportional to $R_{m,i}(r)$ (where, for ease of notation, we denote $R_{m,i}$ as the radial profile $R_{m,\omega}$ for the frequency $\omega = \omega_{m,i}$ mode). 
The radial function $R_{m,\omega}$ of an azimuthal number $m$, angular frequency $\omega$ mode is ultimately obtained by normalising the principal component such that Eq.~(\ref{eqn:R_normalisation}) is satisfied.

The SVD is a principal component analysis (PCA) technique that offers a highly robust and effective way to extract radial functions, with the explained variance ratio (EVR) offering a convenient way to quantify the confidence in the extracted mode. 
In the PCA framework, the EVR is defined as the ratio between the variance captured by the retained singular component(s) and the total variance of the data matrix supplied to the decomposition~\cite{halko2011finding}.
In the present context, the EVR provides a compact, model-independent measure of the degree to which the frequency-radius data admit a low-rank representation. In particular, it quantifies the extent to which the assumption of a separable radial structure underlying the mode remains valid, i.e. that the wave modes are well modelled by Eqs.~(\ref{eqn:general_decomposition}-\ref{eqn:Psi_cylinder}). 
Any deviations from separability arising from noise or additional unresolved structures (i.e. independent modes at the same $m$, and $\omega$) necessarily reduce the EVR value and appear as higher-order singular components, so if these effects are negligible and the EVR remains close to 1 then the $\Tilde{\eta}_m$ frequency band data are well described by a single separable product.

 Repeating the SVD for $\eta_{m,\omega}$ data band-pass filtered around each identified spectral peak in the set $\omega_{m,\{i\}} $, returns a set of extracted mode radial profiles $R_{m,\{i\}} $ at that azimuthal number, and repeating the entire process at each $m$ returns a set of radial profiles $R_{\{m, i\}} $ for all excited modes identified in the data.

\subsection{Mode tracking} \label{Mode_tracking_subsect}

Equipped with a set of mode radial functions and associated frequencies, one can track the corresponding amplitudes throughout time. 
For a fixed value of $m$, let $R_{m,\{i\}}(r)$, $\forall i = 0,1,2,\dots,(N_i-1)$ be the finite set of radial functions, obtained from the mode extraction procedure detailed in Section~\ref{Mode_extraction_subsubsect}. We can express the interface height at azimuthal number $m$ as a superposition of these modes with a small background term $B(t,r)$ that accounts for detection noise and any excited modes missing from the set:
\begin{equation} \label{eqn:etam_expansion_with_background}
    \eta_m(t,r) = \sum_i \xi_{m,i}(t) R_{m,i}(r) + B(t,r).
\end{equation}
We possess 
data of $\eta_m(t,r)$ on a discrete set of radial points $r=r_{\{j\}}$, $\forall j = 0,1,2,\dots,(N_r-1)$. 
At a fixed instant in time $t_k$, Eq.~(\ref{eqn:etam_expansion_with_background}) can be cast as a matrix problem
\begin{equation} \label{eqn:etam_vector_expansion}
    \Vec{E} =  \tensor{R} \Vec{\xi}  + \Vec{B},
\end{equation}
with the matrix elements defined as follows:
\begin{subequations}
\label{eqn:vector_definitions}
\begin{align}
(\Vec{\xi})_{i} &\equiv \xi_{i}(t=t_k), \\
(\Vec{E})_{j} &\equiv  \eta_{m}(t=t_k, r=r_j), \\
(\Vec{B})_{j} &\equiv  B(t=t_k, r=r_j), \\
(\tensor{R})_{j,i}& \equiv  R_{i}(r=r_j).
\end{align}
\end{subequations}
Both the vector $\Vec{E}$, of length $N_r$, and the matrix $\tensor{R}$, of size $N_r\times N_i$, are known. 
The unknown background function $B$ is expected to be small since most of the near-Gaussian detection noise should have been filtered out by the FFT over $\theta$ (this can be verified by inspecting the spectrum $\mathrm{PSD}[\eta_m]$: most of the signal should be within the identified peaks with a comparatively smaller background noise floor below the detection threshold level).
Thus, one can numerically solve for the amplitudes $\Vec{\xi}\approx (\tensor{R}^{\mathrm{T}} \tensor{R} )^{-1} \tensor{R}^{\mathrm{T}} \Vec{E}$ by a least squares minimization. In the presence of unknown background $B$, this is best accomplished with ridge regression (L2-regularised linear regression) as a regularisation procedure, with a regularisation strength parameter selected by cross-validation~\cite{hastie2009elements}.

By performing this procedure at each individual time $t$, one obtains a full evolution of each mode. The advantage of this methodology is that complex amplitudes $\xi_{m,\omega} (t)$ retain the same time resolution as the measurement while still pertaining to a particular oscillation frequency, which is not possible through conventional means of mode tracking through time such as short time Fourier transform~\cite{gabor1946theory, oppenheim1999discrete}, low-pass filtering with demodulation~\cite{oppenheim1999discrete}, or wavelet transform~\cite{torrence1998practical}.

The accuracy of an extracted mode amplitude can be quantified by a signal-to-noise-ratio (SNR) of its spectrum. 
To construct this quantity, one must first calculate the power spectral density of the full time signal $\xi_{m,\omega}$ using the same multitaper periodogram as used previously,
to obtain $\mathrm{PSD}[\xi_{m,\omega}]$ as a function of the continuous frequency variable $\omega^\prime$.
The signal power at a target frequency $\targetomega$, within a band set by the ENBW of the multitaper periodogram calculation, is defined as a functional of the power spectral density:
\begin{equation}
P\!\left[\xi_{m,\omega};\,\targetomega\right]
=
\int_{\targetomega-\mathrm{ENBW}}^{\targetomega+\mathrm{ENBW}}
\mathrm{PSD}[\xi_{m,\omega}](\omega')\,\mathrm{d}\omega'.
\end{equation}
The total power of $\xi_{m,\omega}$ is given by the integral:
\begin{equation}
P_{\mathrm{Tot}}\!\left[\xi_{m,\omega}\right]
=
\int_{-\infty}^{+\infty}
\mathrm{PSD}[\xi_{m,\omega}](\omega')\,\mathrm{d}\omega'.
\end{equation}
Let us consider the amount of ``good signal'' in the extracted mode amplitude $\xi_{m,\omega}$ as the power content at the expected frequency:
\begin{equation} \label{eqn:P_signal_def}
    P_{\mathrm{Sig}}\left[\xi_{m,\omega}\right] \equiv P\!\left[\xi_{m,\omega};\,\targetomega=\omega\right],
\end{equation}
and the unwanted ``noise'' content as all other power content:
\begin{equation}\label{eqn:P_noise_def}
    P_{\mathrm{Noise}}\left[\xi_{m,\omega}\right] \equiv P_{\mathrm{Tot}} - P_{\mathrm{Sig}}\left[\xi_{m,\omega}\right].
\end{equation}
We can therefore define the SNR of an extracted mode amplitude as the ratio of Eq.~(\ref{eqn:P_signal_def}) to Eq.~(\ref{eqn:P_noise_def}):
\begin{equation} \label{eqn:SNR_def}
    \mathrm{SNR}[\xi_{m,\omega}] \equiv P_{\mathrm{Sig}}\left[\xi_{m,\omega}\right] / P_{\mathrm{Noise}}\left[\xi_{m,\omega}\right].
\end{equation}
Because the time series $\xi_{m,\omega}(t)$ is assembled from independent geometric decompositions at each time step, the extraction procedure does not impose any frequency information a priori. Consequently, the resulting signal is not constrained to oscillate at the nominal frequency $\omega$, unlike in cosine-based or band-pass filtering approaches. The SNR therefore provides an independent diagnostic for assessing whether the extracted dynamics exhibit the expected spectral behaviour.

The pipeline behind the EMT protocol is illustrated in Figure~\ref{fig:EMT_overview}
by extracting and tracking the $m =4$, $\omega = 3.1(2\pi)\mathrm{Hz}$ mode with data obtained from the experiment detailed in Sect.~\ref{experimental_setup_subsection}.

\subsection{Generalising to non-cylindrical geometries} \label{Generalisation_subsect}
It should be noted that the EMT methodology can be extended to different vessel geometries.
In principle, one could skip the step of $m$-filtering via FFT and consider the full data $\eta(t,x,y)$ across two arbitrary spatial dimensions $x$ and $y$ instead of $\eta_m$.
Modifying the EMT procedure in this way would enable its application to fluid surfaces within any bounding geometry.
The main conceptual obstacle comes from handling the frequency degeneracy of modes.

In its current form, the EMT method is ill-equipped for handling data containing independent modes oscillating at the same frequency. 
While there is frequency degeneracy in the modal basis of a cylindrical geometry, the issue is circumvented by exploiting the system's axial symmetry.
Since the interface is continuous, the periodic boundary condition $\eta(t,r,\theta)=\eta(t,r,\theta+2\pi)$ is exact.
Fourier transform filtering reduces the (2+1)-dimensional data $\eta(t,r,\theta)$ to (1+1)-dimensional data $\eta_m(t,r)$ for which the modes present at that $m$ value possess unique frequencies. 
Under these conditions, individual radial profiles may be isolated unambiguously within each frequency band.

For more general geometries, frequency-band-limited data, $\Tilde\eta(\omega,x,y)$ containing multiple degenerate modes may still be analysed using principal component analysis. In such cases, several two-dimensional spatial structures $\Psi_{a,\omega}(x,y)$ may be identified via higher-order principal components. However, because PCA separates directions of maximal variance rather than physically meaningful modes, degenerate modes are not uniquely identifiable, and components associated with large-amplitude modes may dominate the signal, reducing the fidelity with which weaker modes are extracted.

Furthermore, the temporal reconstruction stage (tracking amplitudes through time) of EMT is better suited to handling $m$-filtered data $\eta_m(t,r)$ than the full signal $\eta(t,r,\theta)$.
Solving the linear inverse problem of Eq.~(\ref{eqn:etam_vector_expansion}) at each time step is numerically stable since $\tensor{R}$ comprises only a small number of one-dimensional spatial profiles: the set of extracted radial functions at a fixed $m$.
In contrast, a substantially larger set of all of the extracted mode spatial profiles $\Psi_{a,\omega}(x,y)$ would be needed for the equivalent vector equation generalised for 2-dimensional data $\eta(t=t_k,r,\theta)$. 
Consequently, many more degrees of freedom in $\Vec{\xi}$ must be fitted, leading to a high-dimensional and potentially ill-conditioned inverse problem. Such problems are prone to instability and overfitting in the presence of noise.

While the challenges raised here are by no means insurmountable, they require careful consideration when adapting EMT to analyse the data of non-axisymmetric systems.
Either improvements in numerical methodology, or the acquisition of higher-quality datasets may prove sufficient to overcome these difficulties.
Otherwise, they may be circumvented altogether by lifting the frequency degeneracy of the wave mode basis. 
This could be feasible in other experimental systems, whether it is accomplished by a symmetry based dimensional reduction technique, as in the present case, or by considering a system of wave modes that obey a well known dispersion law without frequency degeneracy.

\begin{figure}
    \centering
    \includegraphics[width=3.4in]
{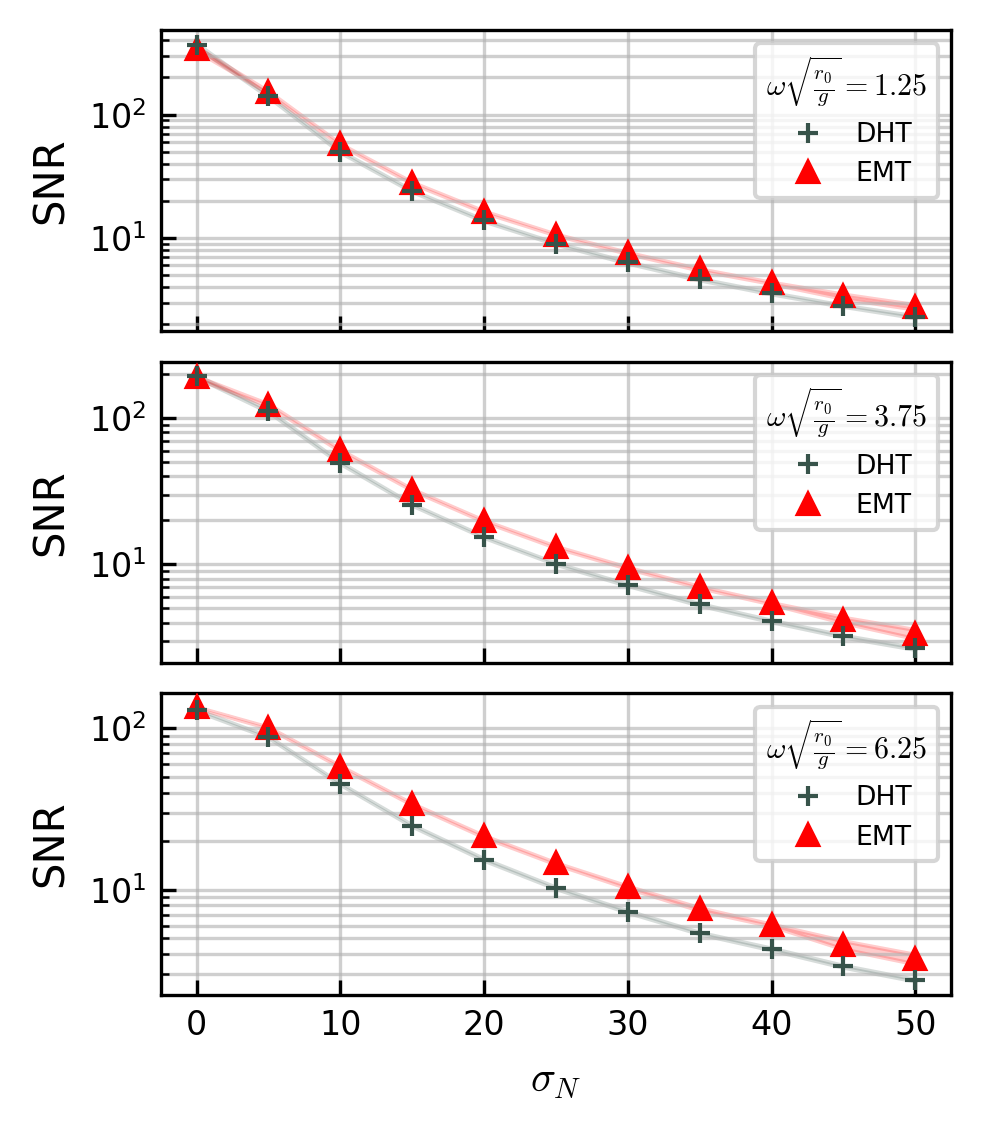}
\caption{\textbf{Tracking accuracy vs noise level} 
Synthetic data of $\eta(t,r,\theta)$ is constructed from synthetic $m=4$ modes at dimensionless frequencies $\omega\sqrt{r_0/g}=1.25, 3.75, 6.25$ and equal envelope amplitudes $|\overline{\xi}_{m,\omega}|/\ro=0.5$. Gaussian noise is added with varying standard deviation $\sigma_N$. The data is processed using both the extracted mode tracking (EMT) and discrete Hankel transform (DHT) methods and, to quantify the precision, signal-to-noise-ratios (SNR) are calculated for extracted modes found at frequencies corresponding to the synthetic modes. For each mode, SNR is plotted against $\sigma_N$ with the 95\% confidence interval from 50 repetitions shaded in the same colour. We see the precision of both methods fall with increasing noise level. We also see the EMT method provides a consistently stronger signal than the DHT method as a result of the ridge regression.
}
\label{fig:SNR_vs_noise}
\end{figure}

\section{Validation} \label{Validation}
\subsection{Numerical benchmarking} \label{EMT_benchmarking}
To validate the experimental mode tracking (EMT) methodology, it is tested on synthetic data. 
Synthetic data on the surface height fluctuations $\eta\Syn$ are prepared
with $N_t$, $N_r$, $N_\theta$ points in time, radius and angle respectively -- these are taken to be $N_t=1000$, $N_r=524$, $N_\theta=64$ unless otherwise specified.
The data is considered to be dimensionless; dimensions of length are rescaled by the vessel radius $r_0$, and dimensions of time by a factor of $\sqrt{r_0/g}$, where $g$ is acceleration due to gravity. The dimensionless time resolution is taken to be $(t_{k+1}-t_k) = 0.1$ and the dimensionless spatial resolution is fixed to $1/N_r$ for radius and $2\pi/N_\theta$ for angle.

The data is generated by summing together a set of wave modes whose radial profiles are chosen to be normalised Bessel functions of the first kind $R_{m,n\omega_{\mathrm{p}}}\Syn = \mathcal{N}_{m,n} J_m (p_{m,n} r/r_0)$ with zeros, $p_{m,n}$, prescribed by a Neumann boundary condition i.e. satisfying $J_m^\prime (p_{m,n})=0$. The normalisation is defined as 
\begin{equation}
    \mathcal{N}_{m,n} \equiv  \langle |J_m (p_{m,n} r/r_0)|^2 \rangle_r^{-1/2},
\end{equation}
so that Eq.~(\ref{eqn:R_normalisation}) is satisfied.
Each wave mode is assigned an amplitude of the form in Eq.~(\ref{eqn:ximn_time_dependence}) with a randomised phase. The values of the dimensionless envelope height and the dimensionless angular frequency were chosen to be comparable to those observed in the experiment presented in Sect.~\ref{Experimental_results_subsection}, they are $\bar\xi_{m,\omega_{\mathrm{p}}}/r_0=0.5$ and $n\times \omega_{\mathrm{p}}\sqrt{r_0/g} = n \times 1.25$ respectively.
Multiple synthetic modes exist at each $m=4,8,12,16$. All tracked modes display the same benchmarking trends though, for the purposes of visualisation, only a few are presented in the following Figures.

Synthetic Bessel modes were selected for the purposes of comparing EMT against an alternative methodology: Discrete Hankel Transform (DHT), for which this synthetic data is the best case scenario. 
The DHT method works by projecting $\eta_m$ onto an eigenbasis of Bessel functions (with known zeros $p_{m,n}$) at each time-step~\cite{guizar2004computation,zhang2023pattern,gregory2024tracking}.

To test the resilience of both EMT and DHT methods against noise, we apply Gaussian noise on top of $\eta\Syn$, representative of that expected from the detection method of our experiment~\footnote{While the direct imaging has Gaussian noise, Fourier transform profilometry introduces length scale dependence to the noise distribution on the reconstructed data. This matters for large length scales such as global shifts to the data but may safely be modelled as Gaussian for local structures such as any mode with $|m|>0$.}. The Gaussian distribution is centred around $z=0$ with a standard deviation $\sigma_N$. Confidence intervals of $95\%$ are calculated from 50 different synthetic data sets at each data point and $\sigma_N$ is varied from 0 to 50. Each dataset is processed using both the EMT and DHT methods. The accuracy of the wave tracking is quantified by the SNR measure defined by Eq.~(\ref{eqn:SNR_def}) and the results are displayed in Fig.~(\ref{fig:SNR_vs_noise}). While increasing the noise level diminishes the SNR of both methods, we see both to be robust, with EMT showing a slight improvement over DHT.

\begin{figure}
    \centering
    \includegraphics[width=3.4in]
{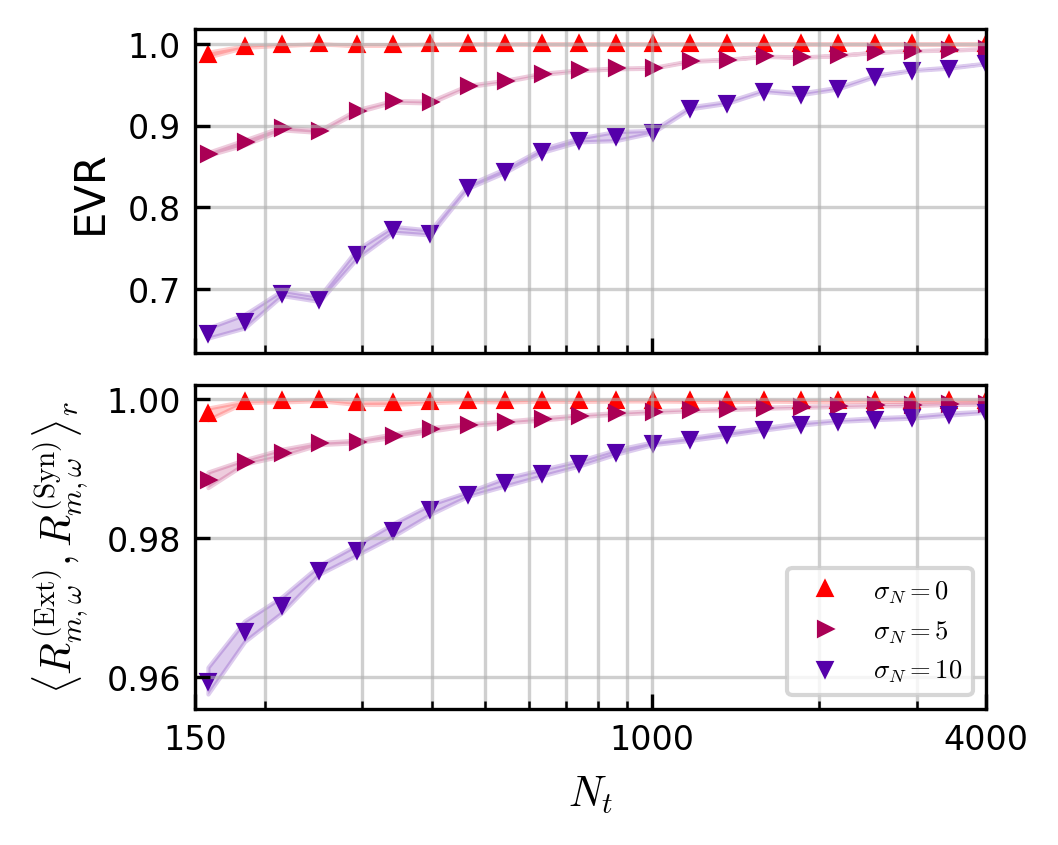}
\caption{\textbf{Radial profile extraction accuracy vs temporal resolution}
The upper panel displays the Explained Variance Ratio (EVR) of the truncated SVD used in the mode extraction procedure, and the lower panel displays the absolute radial correlation between the extracted radial profile $R_{m,\omega}^{(\mathrm{Ext})}$ with that of the corresponding synthetic mode $R_{m,\omega}^{(\mathrm{Syn})}$ present in the data. The former quantity informs us on the success of the SVD and the latter informs us on how accurate the result is. Both quantities are plotted as a function of the number of points in time considered $N_t$ for different noise levels $\sigma_N$. The same trend is observed across both panels.
}
\label{fig:EVR_vs_Nt}
\end{figure}

To quantify the performance of the radial profile extraction, we consider the EVR associated with the truncated SVD for the $\omega\sqrt{r_0/g} = 1.25$ mode as a function of time and imposed noise level $\sigma_N$. 
In addition, we compute the radial correlation between the extracted mode profile $R_{m,\omega}^{(\mathrm{Ext})}$ and the corresponding synthetic profile $R_{m,\omega}^{(\mathrm{Syn})}$ used to generate the data. This correlation is defined as the radial inner product $\langle R_{m,\omega}^{(\mathrm{Exp})}, R_{m,\omega}^{(\mathrm{Syn})} \rangle_r$, and it measures the degree of similarity between the two profiles (independent of their time amplitudes $\xi_{m,\omega}$). The absolute value of this quantity therefore provides a direct measure of how accurately the extraction procedure recovers the true radial structure of the mode present in the data.

Both the EVR and the radial correlation are presented as functions of $N_r$ in Figure~\ref{fig:EVR_vs_Nt}. Increasing the resolution in time, we see an improvement in the SVD computation from the increase in EVR (upper panel), and a better match to the synthetic data from the increase in the radial correlation (lower panel). We see that both quantities decrease with more detection noise $\sigma_N$. The mode extraction procedure will fail entirely if there is insufficient data (less than $N_t\approx150$). 
There is a very close similarity between the trend of the EVR and that of the radial correlation, and so the EVR can be used as a reasonable proxy for quantifying the accuracy of the extracted mode profile. This is convenient when the ``true'' modes present in the data are unknown.

\begin{figure}
    \centering
    \includegraphics[width=3.4in]
{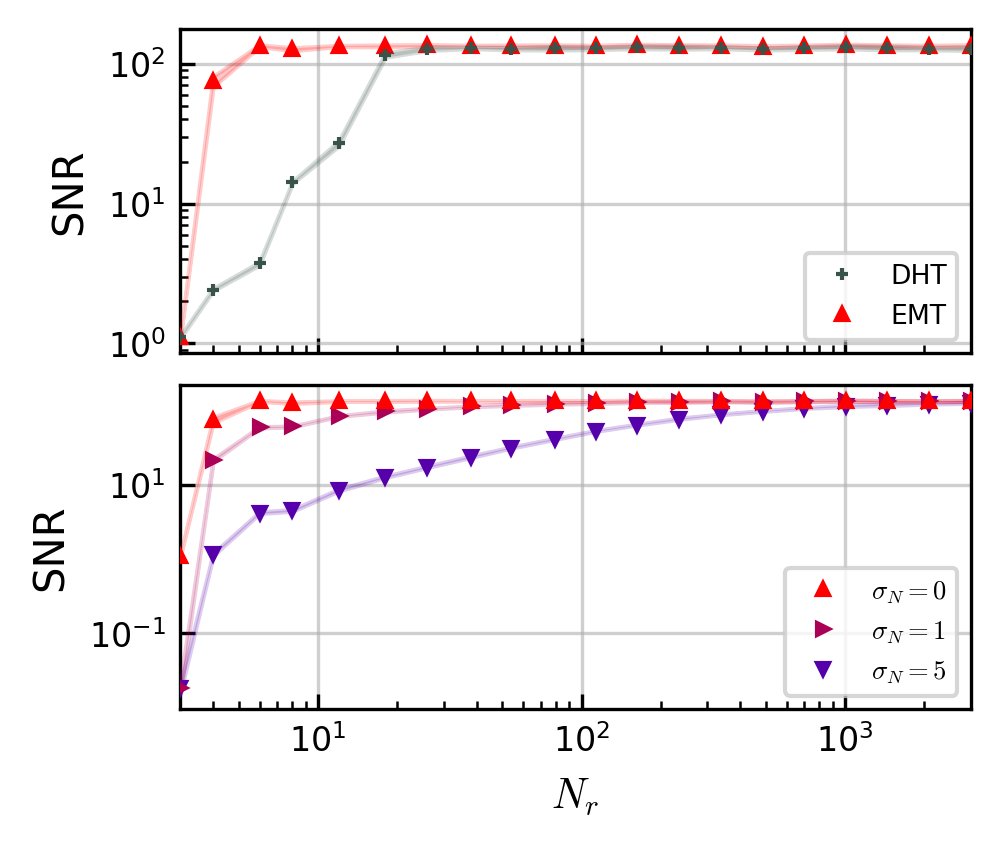}
\caption{\textbf{Tracking accuracy vs spatial resolution}
The SNR of the $m=4$, $\omega\sqrt{r_0/g}=1.25$ mode is presented against the number of data points across radius $N_r$, with a shaded uncertainty from 100 repetitions. The upper panel compares the success of the DHT and EMT methods for synthetic data without noise $\sigma_N=0$. The lower panel presents the EMT method alone at three different noise levels $\sigma_N=0,1,5$.} 
\label{fig:SNR_vs_Nr}
\end{figure}

\begin{figure}
    \centering
    \includegraphics[width=3.4in]
{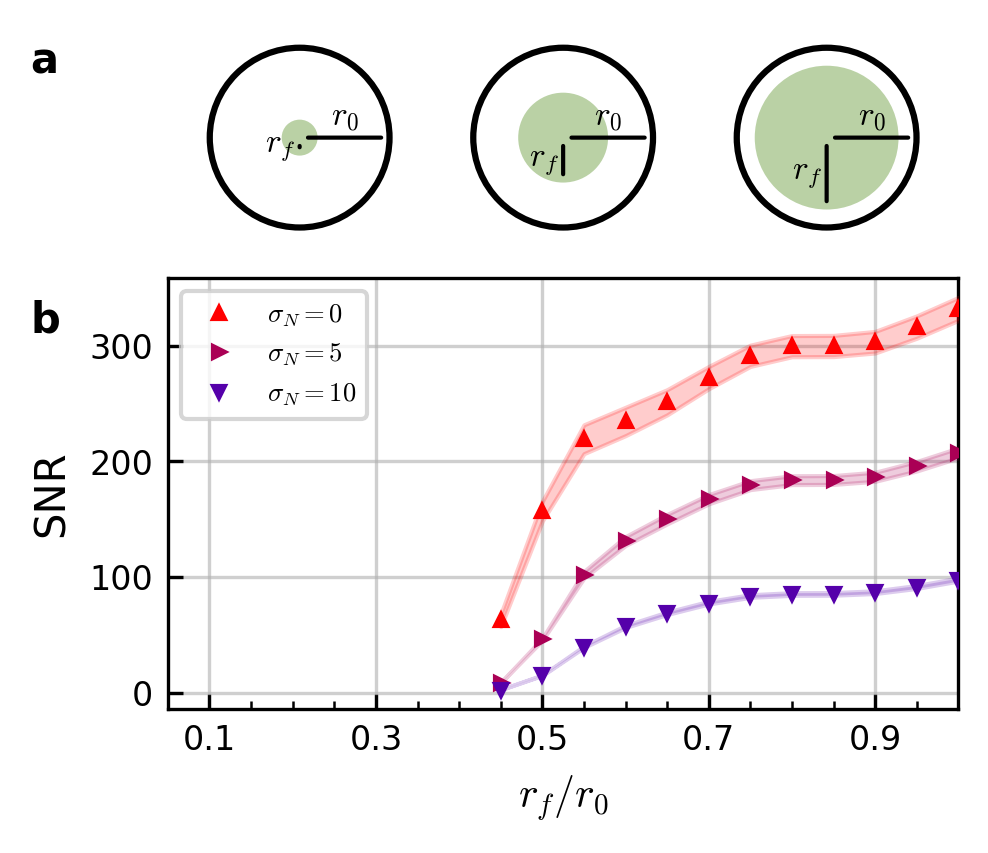}
\caption{\textbf{Tracking accuracy with limited field of view}
Synthetic data is generated and then systematically restricted to circular subregions of radii $r_{f} \leq \ro$, as depicted in panel \textbf{a}. We obtain SNR of the $m=4$, $\omega\sqrt{r_0/g}=1.25$ mode as a function of $r_{f} \leq \ro$ by passing each restricted dataset through the EMT analysis pipeline. In panel \textbf{b} the results are displayed for three different noise levels $\sigma_{N}$ with shaded regions denoting the 95\% confidence interval obtained from 50 repetitions with different synthetic datasets.
}
\label{fig:fov}
\end{figure}

Next, we examine the role of spatial resolution. Limiting the radial resolution does not effect the extraction procedure, but it does lead to poorer tracking results. Figure~\ref{fig:SNR_vs_Nr}, displays the result of varying $N_r$ for the two wave tracking methods (upper panel) and for EMT alone at differing noise levels (lower panel). Without noise, both methods maintain consistent accuracy levels for all radial resolutions except for drastically small number of data points ($N_r<10$). As the noise level increases, we see a continual decrease to lower spatial resolutions, though both methods attain some maximal accuracy above some noise dependent threshold of $N_r$. The DHT method is more limited by radial resolution, since it requires enough points to be able to numerically calculate a Hankel transform; however this requirement is likely to be met for radial resolutions realistic to real world experiments $N_r \sim \mathcal{O}(10^2)$.

Finally we examine how well the EMT method fairs under a limited field of view. Generating synthetic data for $\eta\Syn$ with $N_r=1048$ radial points, we systematically restrict our field of view of this data to circular domains of varying radii $r_{f} \in (0, \ro]$. The SNR of the $m=4$, $\omega\sqrt{r_0/g}=1.25$ mode is again used to quantify the success of the EMT and the results are shown in Figure~\ref{fig:fov}. We see the SNR to decrease with decreasing $r_{f}$. While there is a fall in tracking accuracy once $r_{f}$ drops below $\ro$, it maintains a high SNR and so the method works well with incomplete data. This is a significant advantage over DHT tracking which fails almost immediately on an incomplete domain. We observe the EMT to maintain reasonable quality above around half of the full radius, below which there is a sharp drop off until it fails entirely at low radii. This advantage to EMT is particularly useful for contexts in which the detection measurement does not have access to regions of the spatial domain~\cite{vsvanvcara2024rotating, barroso2025digital}.

\begin{figure*}[ht]
\centering
\includegraphics[width=\linewidth]{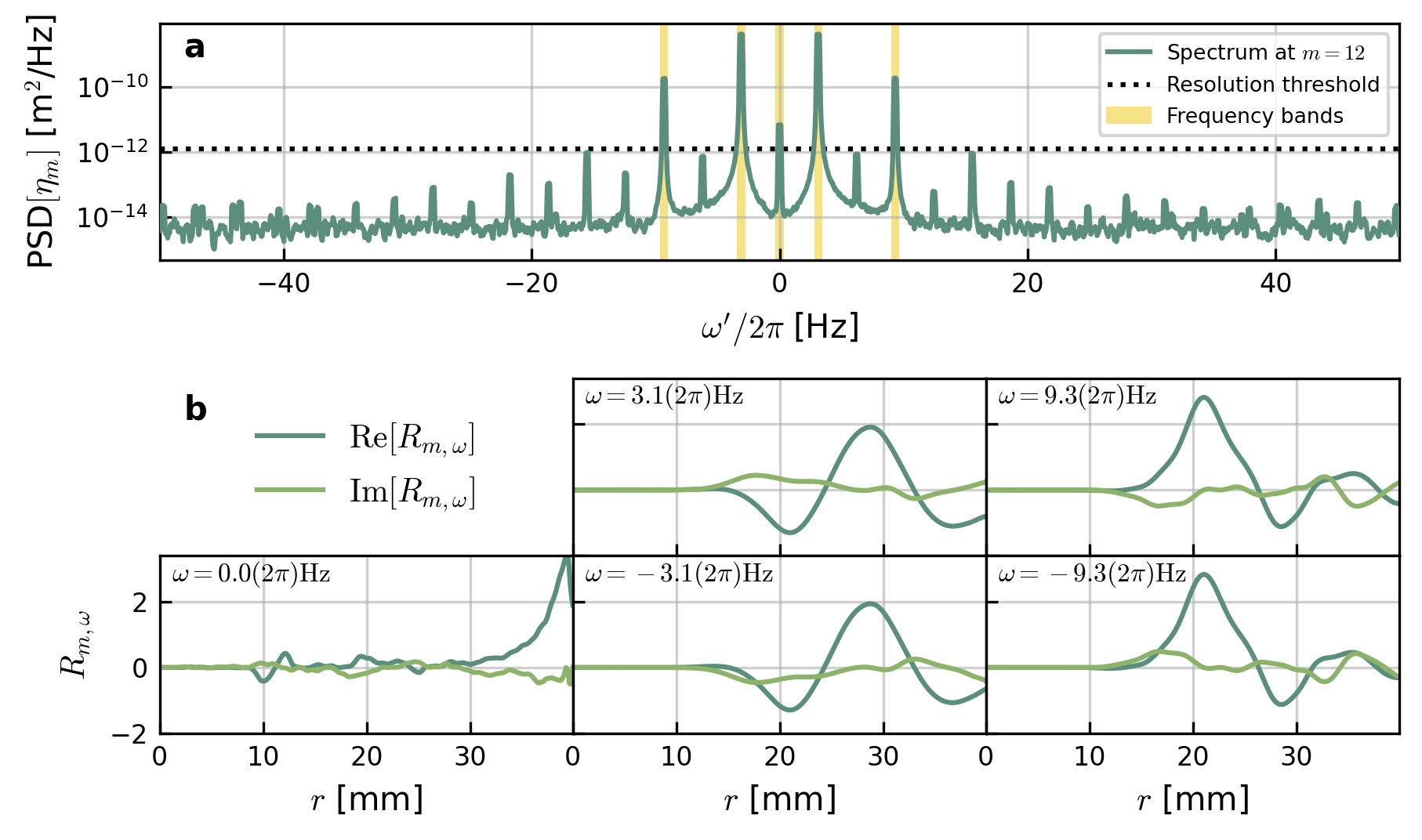}
\caption{\textbf{Radial functions extracted from the frequency spectrum at $m=12$} 
Panel \textbf{a} displays the power spectral density of the interface fluctuations at at $m=12$, as defined by Eq.~(\ref{eqn:PSD_eta_m}), during a time interval $t\in[35.07\mathrm{s},55.03\mathrm{s}]$ during the steady-state. 
Five spectral peaks, with amplitudes above the resolution threshold (denoted by a dotted line), are found and their frequency band is shaded in yellow.
Each peak corresponds to an excited wave mode, and so radial data within each frequency band are used for the purposes of extracting radial functions $R_{m,\omega}$ as detailed in Sect.~\ref{Mode_extraction_subsubsect}.
The radial profiles extracted for each of the five peaks are presented in Panel \textbf{b}.
}
\label{fig:Mode_extraction}
\end{figure*}

\subsection{Experimental results}\label{Experimental_results_subsection}
To demonstrate the mode extraction and tracking procedures of Sect.~\ref{Mode_tracking_section}, they are applied to data from the experimental setup detailed in Sect.~\ref{experimental_setup_subsection}.

Since the azimuthal number $m$ is conserved in wave mixing interactions, significant excitation only occurs at integer multiples of the primary mode's azimuthal number $m=4$~\cite{gregory2024tracking}. For this reason, we only need consider $m=0,4,8,\dots,32$ (the upper limit set by measurement resolution).
For each of these $m$ values, we locate a set of mode frequencies corresponding to peaks in the spectrum $\mathrm{PSD}[\eta_{m}]$ above a threshold amplitude set by the measurement resolution ($\sim \SI{1}{\micro \meter}$).
An example of the PSD at fixed azimuthal number $m=12$ is presented in Fig.~\ref{fig:Mode_extraction}\textbf{a}, with the located spectral peaks highlighted.
For frequencies within each highlighted band, data on $\Tilde{\eta}_m (\omega,r)$ is passed to a principal component analysis algorithm to extract the radial profile at that frequency. Fig.~\ref{fig:Mode_extraction}\textbf{b} displays the radial profiles obtained for each of the five highlighted spectral bands.

\begin{figure}
    \centering
    \includegraphics[width=3.4in]
{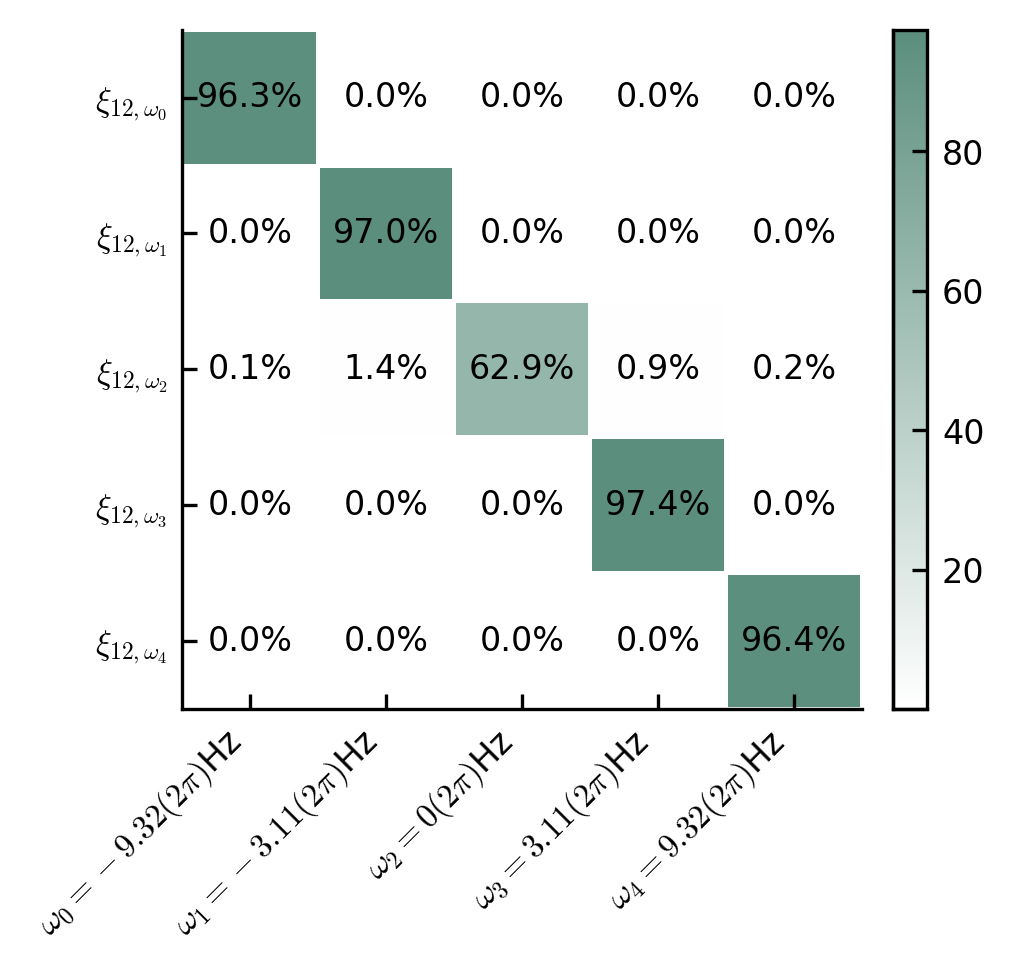}
\caption{\textbf{Confusion matrix} 
The percentage of each $m=12$ mode amplitude's spectrum at different frequencies. Each cell displays a different value $P\!\left[\xi_{m,\omega_a};\,\omega_b\right]/P_{\mathrm{tot}} \times 100 \%$ with $\omega_a$ varying by row and $\omega_b$ varying by column. We see that for the $\omega_a = \omega_0, \omega_1, \omega_3, \omega_4$ modes, $>96\%$ of their spectral content is within the uncertainty band of their expected frequency $\omega_a = \omega_b$, and that there is negligible contribution at the frequencies of other modes $\omega_a \neq \omega_b$. This demonstrates a clear decomposition in which the signal is very well separated into its components with negligible mixing between signals. However, the decomposition of the $\omega_2 = \SI{0}{\hertz}$ mode does not appear as successful.
}
\label{fig:Confusion_matrix_data}
\end{figure}

The extracted mode radial profiles are then used to decompose the fluid interface according to the methodology of Sect.~\ref{Mode_tracking_subsect}. One obtains the full temporal evolution for the amplitude of each extracted mode $\xi_{m,\omega}(t)$.
As before, the precision of the EMT is verified by ensuring that each extracted signal $\xi_{m,\omega}(t)$ has majority of the frequency component at the expected value $\omega$. This can be examined either with the standard SNR measure defined in Eq.~(\ref{eqn:SNR_def}), or 
the total percentage of the signal's power spectrum around a given frequency i.e. $P\!\left[\xi_{m,\omega};\,\targetomega\right]/P_{\mathrm{tot}} \times 100 \%$.
This quantity is presented as a confusion matrix~\cite{stehman1997selecting} for the example tracked $m=12$ modes in Figure~\ref{fig:Confusion_matrix_data}. Most tracked modes display very high precision, with almost the entirety of their frequency content at the expected value ($>96\%$). The exception we see is the zero frequency mode, with $62.9\%$. This lower value would suggest a reasonable but imperfect extraction, likely as a result of the extracted radial function being mixed with those of other frequencies due to aliasing.

A total of 40 excited modes were tracked.
Table~\ref{tab:mode_table} lists the SNR of each mode amplitude found, along with a standard statistical uncertainty calculated from an ensemble of 24 experimental repetitions.
All modes have high EVR values $(>0.95)$ of the PCA used for extracting radial profile, which would suggest that the separability assumption implicit in Eq.~(\ref{eqn:general_decomposition}) is well justified.
Excited modes follow a chequerboard-like pattern across azimuthal number and frequency due to the fact that wave-mixing interactions conserve these quantities~\cite{nazarenko2011wave,Barroso_2023,gregory2024tracking}, hence the azimuthal number and frequency of each mode is an integer multiple of that of the primary mode $m_{\mathrm{p}}=4$, $\omega_{\mathrm{p}}=\pm 3.1(2\pi)\mathrm{Hz}$.

\begin{figure}[h]
    \centering
    \includegraphics[width=3.4in]
{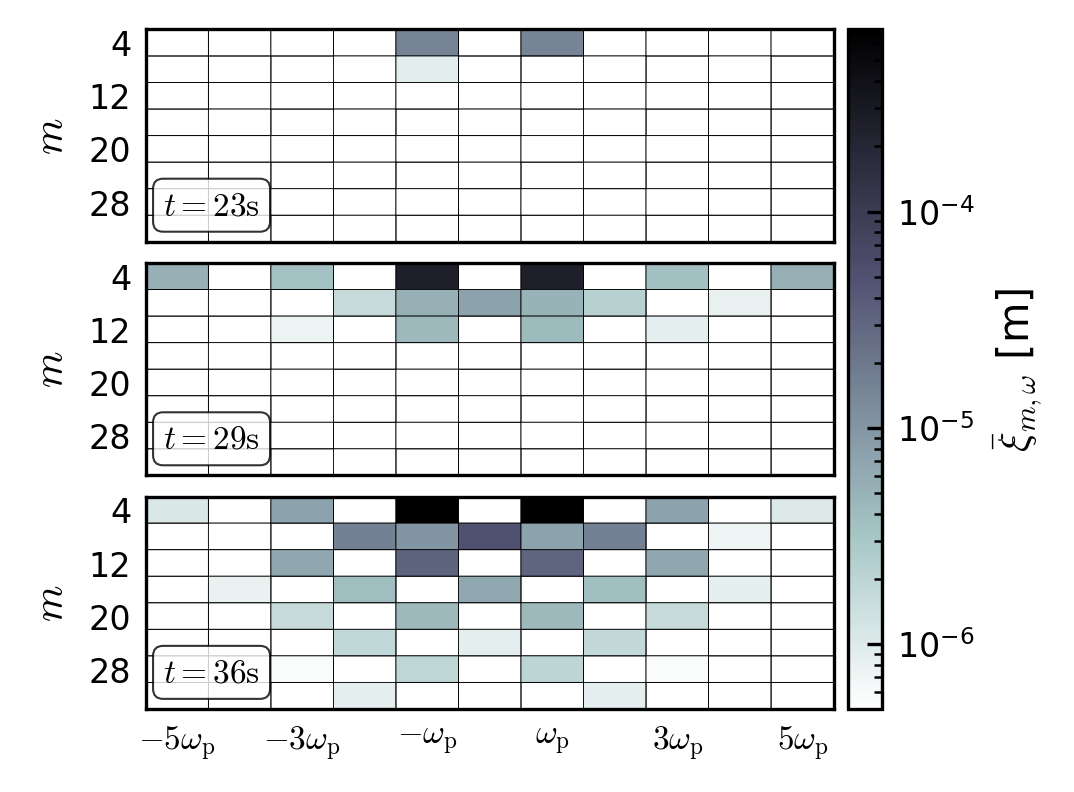}
\caption{\textbf{Time-resolved cascade tree.} 
For all modes considered, the envelope amplitudes $\overline{\xi}_{m,\omega}(t)$ are plotted at three moments in time $t=\SI{23}{\second}, \SI{29}{\second}, \SI{36}{\second}$. 
}
\label{fig:Cascade_tree}
\end{figure}

The chequerboard pattern of excitations can be better seen in Figure ~\ref{fig:Cascade_tree}, which displays the envelope amplitudes of all modes at three different snapshots in time: each a  distinct phase of the system's evolution. The upper panel of Fig.~\ref{fig:Cascade_tree} is taken at an early time $t=\SI{23}{\second}$ when only the primary mode $\xi_{4,\omega_{\mathrm{p}}}$ has been meaningfully excited as a result of parametric resonance. The dynamics at this early stage can be treated linearly. As the amplitude of the primary grows larger, nonlinear interactions become increasingly relevant. These induce the growth of a few other secondary modes, namely $(m,\omega) = (4, \pm3\omega_{\mathrm{p}}),(8, \pm2\omega_{\mathrm{p}}) $ and $(12, \pm\omega_{\mathrm{p}})$ modes. This can be seen on the centre panel of Fig.~\ref{fig:Cascade_tree}\textbf{b}, taken at $t=\SI{29}{\second}$. In turn, these secondary modes interact with each other to source the growth of many more wave modes. Eventually the primary mode saturates due to its quartic order nonlinear self-interaction~\cite{gregory2024tracking}, and all other modes stabilise soon after. At this point a far-from-equilibrium steady state is attained, which persists at late times, as can be seen on the lower panel. By decomposing the data onto a basis of spatial eigenmodes with well defined wavenumber $k$ (but generally mixed frequencies), we have previously shown that this steady state is a direct energy cascade in which the power spectral density obeys a powerlaw scaling behaviour across lengthscales~\cite{gregory2024tracking}. Due to the strong forcing of the system together with its highly discretised basis, the dispersion relation is violated (there is a one to many correspondence between frequency and wavenumber and vice versa) and so scaling behaviour is not observed in the frequency spectrum.
Across the times displayed in Figure~\ref{fig:Cascade_tree}, we observe what is referred to as a cascade tree in which energy is transported from an injection point at $m_p = 4$, $\omega_p=3.1(2\pi)\mathrm{rad s}^{-1}$ out to successively greater values of both $m$ and $\omega$.

Those modes which do not follow the predicted chequerboard pattern are $(m,\omega)=(8,\pm\omega_{\mathrm{p}})$, $(8,\pm3\omega_{\mathrm{p}})$, $(12,0)$, $(16,\pm\omega_{\mathrm{p}})$, and $(20,0)$. Each of these has a low SNR $<7$. A SNR $\sim 1$ indicates that the expected spectral signal Eq.~(\ref{eqn:P_signal_def}) is of the same order as the unwanted noise Eq.~(\ref{eqn:P_noise_def}) and so the extracted time signal is untrustworthy. This is indeed the case for almost all the modes observed outside of the chequerboard, and so one cannot state with certainty that their extracted signal corresponds to a physical wave mode. 
The only exception is the $(m,\omega)=(8,\pm\omega_{\mathrm{p}})$ mode with an SNR $\approx 7$. Its SNR being significantly greater than unity would suggest that the mode is indeed physical, and this is further supported by the fact that it possesses a modestly large steady-state amplitude $|\overline{\xi}_{8,\pm\omega_{\mathrm{p}}}(t \in t^{\mathrm{s}} )| \approx \SI{0.01}{\milli \meter}$.
Upon closer examination of the system's Floquet structure, we notice a $m=8$ mode to also fall within the subharmonic instability band which would explain the presence of $(m,\omega)=(8,\pm\omega_{\mathrm{p}})$ mode: it is a mode unintentionally excited through parametric resonance and competing with $(m,\omega)=(4,\pm\omega_{\mathrm{p}})$.

\begin{figure}
\centering
\includegraphics[width=3.4in]{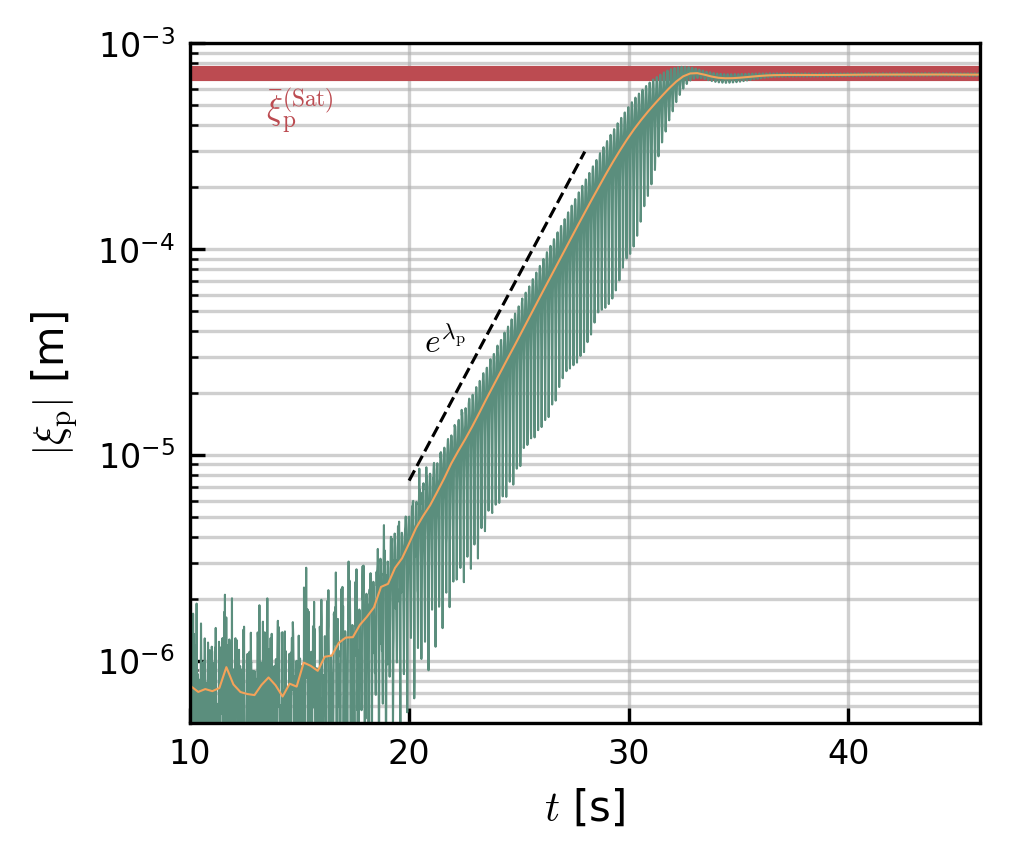}
\caption{
\textbf{Tracked amplitude of primary mode against theoretical prediction.} 
The amplitude of the primary mode $\xi_{\mathrm{p}} = \xi_{m=4,\omega=\omega_{\mathrm{p}}}$ obtained for times $t\in[\SI{10}{\second},\SI{46}{\second}]$. The dark green line denotes the absolute value of the amplitude, $|\xi_{\mathrm{p}}|$, while the orange line denotes the root-mean-square over intervals of $2\pi/\omega_p$, i.e. the envelope amplitude $|\overline{\xi}_{p}|$. The theoretically predicted saturation amplitude $\overline{\xi}_{p}^{(\mathrm{Sat})}=\SI{0.714\pm0.042}{\milli \meter}$ is presented against the data as a red band, and the theoretically predicted exponential growth rate $\lambda_p = \SI{0.46(1)}{\per \second}$ is presented as a dashed line. Both saturation amplitude and growth rate predictions are excellent matches for the experimental values measured from the decomposition: $\overline{\xi}_{4,1}^{(\mathrm{Sat})}=\SI{0.714 \pm 0.006}{\milli \meter}$ and $\lambda_{4,1} = \SI{0.475 \pm 0.003}{\per \second}$ respectively.
}
\label{fig:Mode_frequencies}
\end{figure}

The final demonstration of the EMT's utility we make, is to compare the extracted evolution of the primary mode (the same as used for the demonstration in Fig.~\ref{fig:EMT_overview}) against theoretical prediction. 
From Floquet theory, one can predict the growth rate of the primary mode $\overline{\xi}_{p}\propto e^{\lambda_p t}$~\cite{kumar1994parametric}. For the physical parameters of our experiment, we calculate this growth rate to be $\lambda_p = \SI{0.46(1)}{\per \second}$, which we plot as a dashed line on the top panel of Fig.~\ref{fig:Mode_frequencies}\textbf{a} to show its close match to that seen in the data $\lambda_{4,\omega_\mathrm{p}} = \SI{0.475 \pm 0.003}{\per \second}$. Furthermore, one can use a reduced dynamics approximation scheme to predict the saturation amplitude of the primary mode $\overline{\xi}_{p}^{(\mathrm{Sat})}=\SI{0.714\pm0.042}{\milli \meter}$~\cite{gregory2024tracking}, which also coincides with the observed value of $\overline{\xi}_{4,\omega_\mathrm{p}}^{(\mathrm{Sat})}=\SI{0.714 \pm 0.006}{\milli \meter}$. 
This reduced dynamics approximation scheme follows a procedure developed for the determination of fixed point solutions of Faraday wave dynamics~\cite{miles1984nonlinear} that has been extended to work in any narrow band resonance system such that there exists a primary excitation at significantly larger amplitude than any other~\cite{gregory2024tracking, BHBpaper}.

\section{Conclusion}
Our work presents a novel methodology for the spatio-temporal resolution of gravity-capillary waves in many experimental contexts. By extracting the basis of spatial profiles directly from collected data, one avoids the need to theoretically model the boundary conditions of surface elevation. This is valuable in contexts for which a precise modelling of such boundary conditions is either practically challenging, or reliant on unknown empirical parameters. 
The tracking methodology of geometric fitting at each individual frame provides an excellent time resolution in comparison to other methodologies~\cite{gabor1946theory, oppenheim1999discrete} and, since the frequency information is not used, it comes equipped with a highly efficient way to quantify the success of the tracking through signal-to-noise measures.

We have thoroughly benchmarked the efficacy of the EMT method. Synthetic benchmarking has demonstrated the strong resilience against noise, and the resolution limits are well understood. Furthermore, the ability of EMT to operate on data with restricted spatial domain measurement poses a significant advantage over alternative methods~\cite{zhang2023pattern,gregory2024tracking}, which is of great value to experimental systems lacking this measurement capability.
A Faraday-wave experiment was conducted to further validate the EMT, and highlight the physics that its grants observation access to. 
We report the observation of various nonlinear dynamics, showcasing the potential for studies of nonlinear interaction and wave-turbulence.  
Our work substantially extends the scope of previous fluid–interface studies~\cite{barroso2022primary, gregory2024tracking}, and establishes a framework that will underpin forthcoming investigations of superfluid helium surfaces~\cite{vsvanvcara2024rotating, barroso2025digital}.

\paragraph{\textbf{Acknowledgements}}
We thank members of the Gravity Laboratory for comments and suggestions. 
We acknowledge the support provided by the Leverhulme Research Leadership Award (RL-2019-020),
the Royal Society University Research Fellowship
(UF120112,RF/ERE/210198, RGF/EA/180286, RGF/EA/181015), and partial
support by the Science and Technology Facilities Council (Theory Consolidated Grant  ST/X000672/1), the Science and Technology Facilities Council on Quantum Simulators for Fundamental Physics (ST/T006900/1) as part
of the UKRI Quantum Technologies for Fundamental
Physics programme.

\paragraph{\textbf{Data Availability}}
The data supporting the findings of this study are available from the corresponding author upon reasonable request.

\paragraph{\textbf{Authors contributions}}
S.G. conceived and developed the EMT method, performed all synthetic benchmarking, carried out the data analysis, wrote the manuscript, and co-collected the experimental data. S.S. contributed to the collection of experimental data. V.B. and S.W. designed the experimental setup. V.B., A.A. and S.W. supervised the project and provided feedback throughout the research and manuscript preparation.

\paragraph{\textbf{Competing interests}} The authors declare no competing interests.

\paragraph{\textbf{Correspondence and requests for materials}} should be addressed to Sean Gregory.

\bibliography{Biblio_ArXiv}

\begin{table*}
\caption{
\textbf{SNR of all tracked modes.} The SNR of all 40 modes tracked in the experiment are listed (to 2 decimal point precision) against the azimuthal number $m$ (rows) and oscillation frequency $\omega$ (columns). Empty cells correspond to $(m,\omega)$ values at which no excited mode was detected above the threshold set by measurement resolution. Each SNR value is presented with a standard statistical uncertainty calculated from an ensemble of 24 experimental repetitions. The EVR of the radial profile extraction is $>0.95$ for all modes presented.
}
\label{tab:mode_table}
\centering
\resizebox{\linewidth}{!}{%
\begin{tabular}{r||c|c|c|c|c|c|c|c|c}
\hline\hline
$\omega = $ &
$-4\omega_{\mathrm{p}}$ &
$-3\omega_{\mathrm{p}}$ &
$-2\omega_{\mathrm{p}}$ & 
$-\omega_{\mathrm{p}}$ & 
$0$ & 
$\omega_{\mathrm{p}}$ & 
$2\omega_{\mathrm{p}}$ & 
$3\omega_{\mathrm{p}}$ & 
$4\omega_{\mathrm{p}}$
\\
\hline\hline
m=4 &  
& $6.08 \pm 0.79$ &  & $140.76 \pm 0.99$ &  & $140.82 \pm 0.96$ &  & $6.69 \pm 0.92$ &  \\
m=8 & 
$17.55 \pm 2.03$ & $0.35 \pm 0.04$ & $242.86 \pm 8.15$ & $6.30 \pm 0.85$ & $488.87 \pm 11.88$ & $6.92 \pm 0.96$ & $225.73 \pm 10.98$ & $0.24 \pm 0.03$ & $16.83 \pm 2.18$ \\
m=12 &
& $33.40 \pm 3.46$ &  & $104.39 \pm 4.57$ & $1.24 \pm 0.15$ & $105.40 \pm 4.81$ &  & $34.42 \pm 3.42$ &  \\
m=16 & 
$4.60 \pm 0.67$ &  & $42.12 \pm 5.93$ & $1.33 \pm 0.22$ & $77.62 \pm 10.89$ & $1.60 \pm 0.39$ & $38.71 \pm 4.98$ &  & $5.40 \pm 0.75$ \\
m=20 & 
& $19.33 \pm 2.04$ &  & $72.41 \pm 5.06$ & $2.03 \pm 0.19$ & $67.30 \pm 4.70$ &  & $20.25 \pm 1.83$ &  \\
m=24 & 
&  & $30.38 \pm 1.34$ &  & $4.62 \pm 0.59$ &  & $30.04 \pm 1.56$ &  &  \\
m=28 & 
&  $6.35 \pm 0.68$ &  & $36.92 \pm 3.33$ &  & $39.63 \pm 3.55$ &  & $6.92 \pm 0.88$ &  \\
m=32 & 
&  &$12.18 \pm 0.94$ &  & $10.52 \pm 2.10$ &  & $12.07 \pm 0.95$ &  &  \\
\hline
\end{tabular}
}
\end{table*}

\end{document}